\documentclass{article}
\usepackage{arxiv}
\usepackage{amsmath,amsfonts}
\usepackage{algorithmic}
\usepackage{algorithm}
\usepackage{array}
\usepackage[caption=false,font=normalsize,labelfont=sf,textfont=sf]{subfig}
\usepackage{textcomp}
\usepackage{stfloats}
\usepackage{url}
\usepackage{verbatim}
\usepackage{multirow}
\usepackage{multicol}
\usepackage{graphicx}
\usepackage{pgfplots}
\usepackage{cite}
\definecolor{r4}{HTML}{CE2029}
\definecolor{s3}{HTML}{8DB600}
\usepackage[utf8]{inputenc} 
\usepackage[T1]{fontenc}    
\usepackage{hyperref}       
\usepackage{url}            
\usepackage{booktabs}       
\usepackage{amsfonts}       
\usepackage{nicefrac}       
\usepackage{microtype}      
\usepackage{lipsum}
\usepackage{graphicx}
\graphicspath{ {./images/} }

\title{Cloud Computing Review: A Decade of Research}

\author{
 Smruti Rekha Swain\\
  Department of Computer Science and Engineering\\
  SRM University\\
  Delhi-NCR, 131029 \\
  \texttt{smruti.sai90@gmail.com}} 

\begin{document}
\maketitle
\begin{abstract}
The popularity and rapid development of Cloud
Computing in recent years has led to a vast number of publications capturing the accumulated knowledge in this field. Due to the interdisciplinary nature and significant relevance of cloud computing research, it has become increasingly challenging
to comprehend the overall structure and progress of this field without employing analytical methods. While the evaluation of scientific research has a long tradition in many fields, we have identified a lack of comprehensive scientometric studies
specifically focused on cloud computing. This study applies scientometric techniques to empirically examine cloud computing research’s evolution and current state from a macroscopic perspective. We employed the CiteSpace tool for visual analysis,
exploring topics related to cloud computing by retrieving papers published between 2014 and 2023 from the Web of Science Core database. Our approach involved constructing collaboration networks
among authors, institutions, and countries to pinpoint the most prolific contributors in each category. Through the analysis of core journal distributions via journal co-citations, document
co-citation networks, and clustering analysis, we uncovered the underlying research topics and knowledge structure. The results of this study enhance our understanding of patterns, trends, and
other critical factors, offering a foundation for guiding research activities, sharing knowledge, and fostering collaboration in the field of cloud computing research.
\end{abstract}
\section{Introduction}
Cloud computing is an emerging technique for enabling
ubiquitous, convenient, and on-demand access to
a shared pool of customized computing resources \cite{1,2,3,4,5,6,7,8,9}. The rapid development of cloud computing can be attributed to its interdisciplinary nature, along with the various technical and non-technical potentials and challenges it encompasses \cite{10,11,12,13,14,15,16,17,18,19,20}. As the
number of publications on cloud computing continues to grow significantly, it becomes increasingly vital to assess the current state and trajectory of research in this field. Quantitative studies that measure and analyze scientific activities constitute a type of research commonly known as scientometrics. By offering a meta-perspective on a research field scientometric studies aid in the advancement and enhancement of an academic discipline, serving as a critical foundation for defining and deliberating future research agendas \cite{31,32,33,34,35,36,37,38,39,40}. With the assumption that scientific activities are mirrored in scientific publications, scientometric studies utilize empirical measures to analyze the
scientific output of a particular field. This analysis aims to enhance the understanding of the dynamics and structure of its development \cite{21,22,23,24,25,26,27,28,29,30}. As a result, it becomes possible to explore the
body of publications extensively, allowing for the observation of various aspects such as citation patterns, the number and types of citations, the number and structure of authors, and so on. Furthermore, a scientometric study provides insights
into research activities in general, including aspects related to knowledge sharing, research quality, socio-organizational structures, influential countries/affiliations/authors, the development of key topics, structural changes, and the economic impact of research.

\par
Considering these facts, it is indeed surprising that there hasn't been much focus on the scientometric analysis of cloud computing research, especially concerning a comprehensive scientometric study of the field. The authors of \cite{sivakumaren2012growth} delve into 510 publications of cloud computing, sourced from the Web of Science (WoS) database spanning the years 2001-2010. Their analysis focuses on author productivity and the contributions of different countries, achieved by examining the number of publications aggregated by WoS. In \cite{bai2011scientometric}, the authors examine 89 journal papers related to cloud computing research in China, covering the period from 1993 to 2010. Utilizing data from the Chinese Journal Full-text Database (CNKI), they investigate the distribution of the number of journal papers, authors, subjects, and funded papers \cite{41,42,43,44,45,46,47,48,49,50}. In \cite{wang2013research}, scientometric methods are employed to analyze the research advancement in cloud security from 2008 to 2011 in China. The authors examine 103 journal articles from 76 journals provided by CNKI. They analyze the types of contributing affiliations and identify key topics exclusively focusing on cloud security. In general, these studies lack important insights, such as providing an overview of current research topics and trends, citation patterns, and top publications. The authors in \cite{heilig2014scientometric} conducted a Scientometric Analysis of Cloud Computing Literature from 2008 to 2013, aiming to provide insights into publication patterns, research impact, and research productivity. \\
\par
The implications of these studies are somewhat restricted due to the relatively limited number of publications analyzed and the narrow focus of their research. Furthermore, these studies predominantly utilize basic counting measures to examine the literature \cite{51,52,53,54,55,56,57,58,59,60}. Without incorporating specialized scientometric methodologies, such as diverse approaches to assess author productivity or algorithms for keyword cluster analysis, it becomes difficult to produce innovative insights \cite{61,62,63,64,65,66,67,68,69,70}. As a result, the primary objective of the study is to provide empirical and relevant findings while offering a more comprehensive review of all aspects of cloud computing research.  
\begin{figure*}[!htbp]
  \centering
\includegraphics[scale=.89]{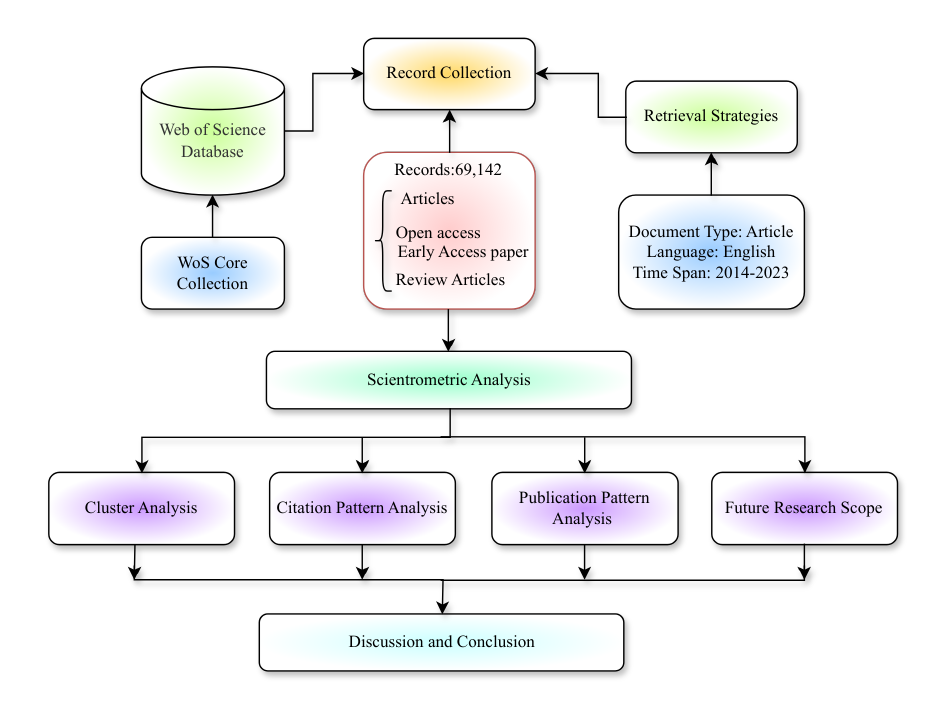}
\caption{Research Methodology}
\label{fig}
\end{figure*}
In this paper, we introduce a comprehensive scientometric study that empirically investigates publications of cloud computing, encompassing data from the Web of Science database spanning from 2014 to 2023. In total, we analyze 69,142 publications, marking the first-ever scientometric study to scrutinize such a vast number of peer-reviewed publications. This study offers comprehensive insights into publishing patterns, such as the countries contributing and the distribution of outlets. Additionally, we analyze keyword clusters to identify widely discussed topics and their interrelationships. In addition, we provide information on the citation patterns of outlets, publications, affiliations, and authors. Lastly, we investigate the research productivity of affiliations and scientists in the field of cloud computing.
In pursuit of this goal, the present study aims to address the following inquiries:
\begin{enumerate}
    \item What key areas of research are prominent within cloud computing?
    \item What is the current academic state of existing cloud computing research in terms of publication and citation patterns over the years, geographical distribution, prominent journals, and productive nations in individual research areas from 2014 to 2023?
    \item What are the implications of the current study, future research directions, and key challenges in the field of cloud computing?
\end{enumerate}
The remainder of this paper is organized as follows: Section II delineates the approach employed for data collection and tool selection for processing the data. Additionally, it identifies the research domains of cloud computing through document cocitation cluster analysis. Section III delves into the growth of publications, their geographical distribution, and country collaboration. Section IV examines the citation patterns of publications and the most prominent journals in various cloud computing domains. Section V sets the stage for identifying potential research directions. while Section VI presents the conclusion, implications, and future directions of the study.
\section{Methodology}
The foundation for a scientometric analysis of a specific research area is established through the collection of relevant publications and citations. As indicated, this study aims to cover a substantial portion of peer-reviewed cloud computing articles published in the past ten years.
\subsection{Framework}
The present study utilizes a scientometric approach to investigate evolutionary paradigms and explore future research directions in the area of cloud computing. Fig. 2 illustrates the framework demonstrating the methodology employed in the current article to visualize and analyze the literature on cloud computing. The analysis steps are outlined as follows:
\begin{enumerate}
    \item The authors gathered literature from the WoS database using diverse restriction criteria. The records were downloaded and stored in a format compatible with CiteSpace.
    \item The co-citation network within the cloud computing literature is created using the CiteSpace visualization tool to explore various domains comprehensively. Additionally, the entire dataset is divided into domains using WoS subqueries.
    \item A thorough scientometric analysis covers all categories of cloud computing. By scrutinizing cloud computing literature within distinct domains, the study unveils trends in publication growth, and collaboration among countries, and identifies high-yield nations. Analyzing citation patterns illuminates distribution and pinpoints the most influential journals within each domain.
    \item An analysis of cocitation clusters, derived from references, identified emerging research frontiers within specific domains for future investigation.
\end{enumerate}
\subsection{Database and Tools Selection}
This research article delves into analyzing literature on Cloud computing within the field of computer science. Several databases are available for scientific literature, such as Scopus, Web of Science (WoS), IEEE Explore, Google Scholar, and Science Direct \cite{harzing2016google} \cite{71,72,73,74,75,76,77,78,79,80}. In this study, however, WoS is selected for its repository of high-impact scientific journals, housing papers of superior quality. It has vast content coverage from 1990 and contains higher impact scientific literature as compared to Scopus \cite{chadegani2013comparison} \cite{81,82,83,84,85,86,87,88,89,90}. Additionally, WoS facilitates on-site analysis and visualization of structured publication data, presenting information through bar charts and treemap charts across diverse parameters such as authors, publication year, document types, WoS categories, affiliations, publication titles, publishers, editors, research areas, country/region, and languages.
\par 
Moreover, manually processing extensive amounts of bibliometric data is both costly and labor-intensive. A visualization tool is necessary for conducting scientometric analysis on structured input and for extracting hidden information from the data \cite{chen2019visualizing}. Several data visualization tools are available for use, such as CiteSpace \cite{chen2012emerging}, Gephi \cite{jacomy2014forceatlas2}, VOSviewer \cite{van2010software}, SciMAT \cite{cobo2012scimat}, UCInet \cite{johnson1987ucinet}, and more. The authors utilized the CiteSpace V.6.2 R6 for data processing. It is a paid software providing distinctive features for evaluative analysis via network visualization \cite{chen2006citespace} \cite{91,92,93,94,95,96,97,98,99,100}. Furthermore, CiteSpace software efficiently organizes results by generating clusters to highlight research areas and overall trends \cite{fang2018climate}. Furthermore, CiteSpace V.6.2 R6 offers various parameters such as citation burstiness, silhouette score, modularity, and centrality to validate the results obtained through visualization.
\subsection{Data Search Strategy}
The Preferred Reporting Items for Systematic Reviews and Meta-Analysis (PRISMA) approach, proposed by Liberati et al. \cite{liberati2009prisma}, has emerged as a widely popular and highly suitable method for assessing scientific data. The current study employed PRISMA to gather the most pertinent scientific literature on cloud computing. As illustrated in Fig. 3, the initial stage retrieved a significant count of 69,142 records from the WoS database using the following query: ALL (Cloud Computing). 
Moreover, during the data selection process, emphasis was given to extracting records from the title, abstract, and keywords fields, resulting in a compilation of 40,637 articles. Through the screening stage, 7,686 records were eliminated by filtering results according to DOCUMENT TYPES ("Article"), LANGUAGES("English"), and PUBLICATION YEARS ("2014-2023").  Eligibility assessment involved removing papers contributing to the computer science subject area, accomplished by applying filters such as WoS CATEGORIES ("Computer Science Information Systems") and research areas ("Computer Science"). Once the eligibility stage was completed, the CiteSpace tool was employed to remove duplicate records from the pool of 12,738 articles. Following the retrieval strategy outlined above, a total of 12,638 records were gathered for scientometric analysis. The search was then refined by manually screening the abstracts of full-text articles to establish the groundwork for an in-depth review process. This thorough process ensures that the resulting collection is free from duplication, thereby strengthening the overall accuracy and reliability of the study.
\begin{figure*}[htbp]
  \centering
\includegraphics[scale=.89]{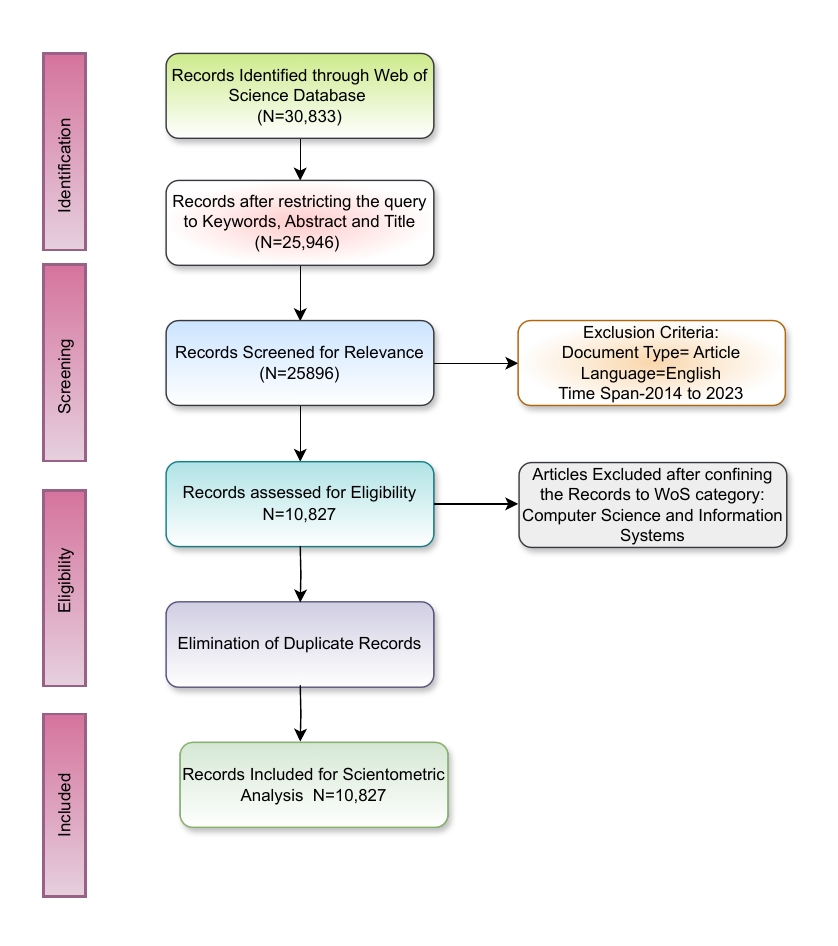}
\caption{Prisma Flow Diagram for Record Selection}
\label{fig}
\end{figure*}
 \subsection{Identification of Research Domains}
The co-citation clusters formed by the links between several publications highlight similar research axes and key issues. By analyzing the degree of correlation between documents, clustering analysis can categorize a large number of research articles into distinct units, enabling the identification of various research domains within emerging technology. 
Therefore, CC research domains are identified by creating document co-citation clusters from records obtained in the WoS database, utilizing the CiteSpace visualization tool. Fig. 3 illustrates the clusters, with each cluster represented by a different color. The number following the \# symbol denotes the cluster ID, and the subsequent phrase indicates the cluster tag. The log-likelihood ratio (LLR) algorithm is employed for cluster generation due to its ability to produce high-quality clusters with high intra-class similarity \cite{ma2022visualization}. The quality of clusters is determined by the Modularity (Q) and mean silhouette score (S). The "Q" value assesses the ease with which a network can be divided into multiple modules, while the "S" value measures the homogeneity of clusters. A network achieves a high level of overall clarity when both S and Q values are close to 1.0. The document co-citation analysis yielded clusters with a Q value of 0.6876 and an S value of 0.8686, indicating the high reliability of the clusters. Table I presents the top 10 most significant clusters of co-cited documents, including the cluster ID, size, silhouette score, mean year, and cluster tag. Each significant cluster comprises the most frequently cited articles, with the cluster tag indicating the main research topic in CC. Based on the cluster tag, all the obtained clusters are broadly classified into five domains: Mobile Edge Computing (MEC) (\#0, \#6, \#12), Fog Computing (FC) (\#1, \#4), Secure Cloud Storage (SCS) (\#2, \#3), Workflow Scheduling (WS) (\#8, \#9) and Federated Learning (FL) (\#5).
Furthermore, to explore the CC domains identified through document co-citation clustering, the entire dataset is dissected using subqueries in the WoS database. Table II displays the WoS query used to segregate data into corresponding domains and the number of records obtained in each domain. These subqueries include the most relevant and frequent keywords to accurately divide the entire dataset into distinct domains. Subsequently, the record contents are downloaded and saved in plain text file format for further evaluation.
\begin{figure*}[!htbp]
  \centering
\includegraphics[width=1.29\textwidth]{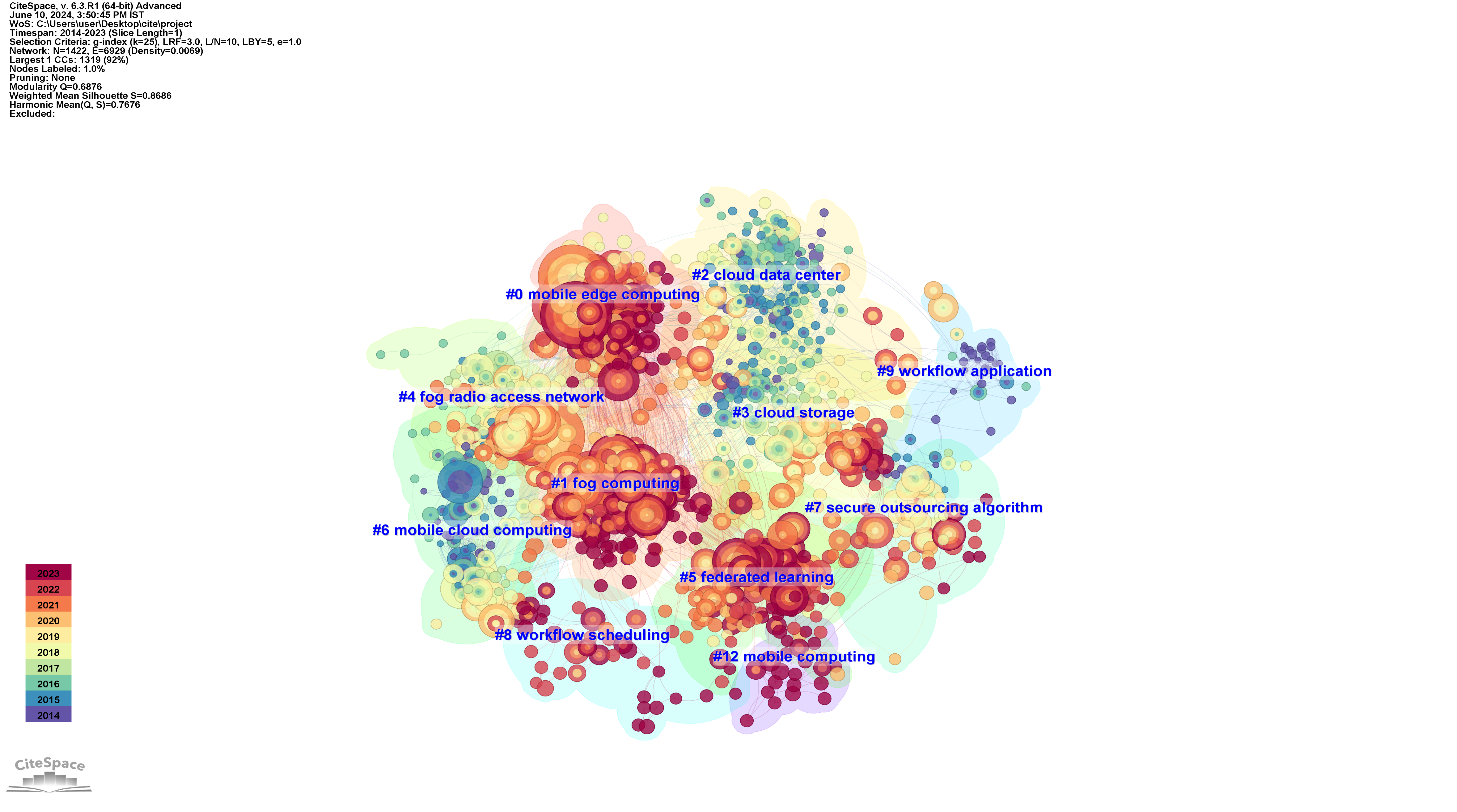}
\caption{Document Co-citation Analysis}
\label{fig}
\end{figure*}
\begin{table} [!htbp]
    \centering
    \caption{ OVERALL CLUSTERS SUMMARY OF Cloud Computing}
    \resizebox{\columnwidth}{!}{
    \begin{tabular}{ |c|c|c|c|c| } 
    \hline
    Cluster Id & Size & Silhouette Score & Mean Year &  Cluster Tag (LLR)\\ [0.5 ex]
    \hline 
    0 & 220 & 0.796 & 2018 & Mobile Edge Computing\\ \hline
    1 & 187 & 0.782 & 2018 & Fog Computing \\ \hline
    2 & 152 & 0.831  & 2012 & Cloud Data Center\\ \hline
    3  & 145 & 0.891 & 2014  & Cloud Storage \\ \hline
    4  & 137 & 0.9 & 2015 & Fog Radio Access Network\\ \hline
    5  & 135 & 0.91 & 2018 & Federated Learning\\ \hline
    6  & 109 & 0.889 & 2012 & Mobile Cloud Computing\\ \hline
    7  & 53 & 0.957 & 2016 & Secure Outsourcing Algorithm\\ \hline
    8  & 44 &  0.969& 2018 & Workflow Scheduling\\ \hline
    9  & 43 & 0.941 & 2010 & Workflow Application\\ \hline
    12  & 23 & 0.981 & 2019 & Mobile Computing\\ \hline
 
    \end{tabular}}
   
    \label {table:1}
    
    \end{table}
\begin{table} [!htbp]
    \centering
    \caption{ Wos Sub Queries for Each Domain}
    \resizebox{\columnwidth}{!}{
    \begin{tabular}{ |c|c|c| } 
    \hline
    Domain & Query & Total Records \\ [0.5 ex]
    \hline 
    Mobile Edge Computing & mobile edge computing or mobile computing  & 3540 \\ \hline
    Fog Computing & fog computing & 2089 \\ \hline
    Secure Cloud Storage & cloud storage  & 1366 \\ \hline
    Workflow Scheduling & workflow scheduling & 1407 \\ \hline
    Federated Learning & federated learning & 4133 \\ \hline 
    \end{tabular}}
   
    \label {table:1}
    
    \end{table}
\subsection{Data Analysis}
Upon importing the data into the science mapping software CiteSpace, two key techniques used for scientometric analysis and visualization are co-word analysis and co-citation analysis. Co-word analysis examines the co-occurrence of keywords or key terms in the publications, identifying relationships between concepts, topics, or research themes based on the frequencies and patterns of keyword co-occurrences. In contrast, co-citation analysis focuses on the citation relationships between publications, journals, and authors, identifying those frequently cited together, which indicates a conceptual or intellectual link. Co-citation analysis helps uncover emerging trends, research fronts, and the evolution of scientific disciplines over time. 
The insights gained from these analyses in CiteSpace can help study the following:
\begin{enumerate}
    \item \textbf{Temporal Distribution of Publications:}
    In the context of scientometric analysis, the temporal distribution of publications refers to how scholarly works have been chronologically spread and evolved. This study analyzes the two types of distribution of publications: Annual distribution of publications and Geographical Distribution of Publications
   
   \item \textbf{Co-Citation Analysis:} 
    A co-citation relationship occurs when two or more articles, authors, or journals appear in a third bibliography [31]. CiteSpace facilitates co-citation analysis by constructing professional structures or maps, monitoring the developmental trends of scientific fields, and evaluating the interrelationships between different professions. This evaluation typically encompasses three types of co-citation analysis: authors co-citation analysis, journals co-citation analysis, and documents co-citation analysis [32], [33].
    \item  \textbf{Co-Word Analysis: }
    Co-word analysis operates under the premise that when two words or concepts frequently appear together in the same document, they are likely related in terms of meaning, context, or thematic association. This method begins by identifying pairs of words or concepts that co-occur within the document, such as in the title, abstract, or keywords.
\end{enumerate}
\section{Visualization Scientometric Analysis}
In this section, the authors perform a scientometric and systematic analysis of the literature in the field of cloud computing. The objective is to uncover bibliometric aspects and extract knowledge from the available publications.

\subsection{Temporal Distribution of Publications}
The analysis of publications reveals the shifting patterns of developmental trends over the years, illustrating the progress of literature in cloud computing. Fig. 5 illustrates the yearly development of CC literature from 2014 to 2023. The trend line for total publications indicates the number of publications each year, while the cumulative frequency depicts the overall growth rate of the literature. The cumulative number of papers per publication year shows a consistent growth in the scientific literature over the decade, illustrating the revolutionary paradigm of the field. The analysis reveals a steady growth in publications until 2017, followed by a sharp increase after 2018. Publications doubled from 2018 (1056) to 2021 (2063), indicating a significant spike in CC-related research over the last past five years.
\begin{table*}[!htbp]
\centering
\resizebox{\textwidth}{!}{
\begin{tabular}{|p{3cm}|c|c|c|c|c|c|c|c|c|c|c|c|c|c|c|c|c|c|c|c|}
\hline
\multirow{2}{*}{Domain}& \multicolumn{2}{|c|}{2014} & \multicolumn{2}{|c|}{2015} & \multicolumn{2}{|c|}{2016} & \multicolumn{2}{|c|}{2017} & \multicolumn{2}{|c|}{2018} & \multicolumn{2}{|c|}{2019} & \multicolumn{2}{|c|}{2020} & \multicolumn{2}{|c|}{2021} & \multicolumn{2}{|c|}{2022} &  \multicolumn{2}{|c|}{2023}  \\  \cline{2-21}

  & TP         & TP\%        & TP         & TP\%        & TP         & TP\%        & TP         & TP\%        & TP         & TP\%        & TP         & TP\%        & TP         & TP\%        & TP        & TP\%        & TP         & TP\%        & TP & TP\% \\  \hline

Mobile Edge Computing &4 &0.113 &7&0.198 &11 &0.311 &64 & 1.80 & 153 & 4.32 &  376 &  10.62 & 526 & 14.85 & 676 &   19.09 & 710  & 20.05 & 720  & 20.33   \\ \hline
Fog Computing &1 & 0.048 & 3 & 0.144 & 15 & 0.718 & 88 &  4.213 & 195 & 9.335 & 283 & 13.547 & 403 & 19.29 &  398 &   19.052 & 426 & 20.393 & 13.260 & \\ \hline
Secure Cloud Storage &27 &  1.977 & 34 & 2.48 & 49 & 3.58& 90 &6.58 &124 & 9.07 &170 & 12.44 & 222 & 16.25 & 222 & 16.69 &  444& 17.27&555& 13.61 \\ \hline
Work Flow Scheduling & 67 & 4.76 & 69 & 4.90 & 94 & 6.68 &  108 & 7.676 & 96 & 6.823 & 150 & 10.66 & 203 & 14.428 & 227 & 16.134 & 207 & 14.712 &186 & 13.220    \\ \hline
Federated Learning & 7 & 0.18 & 10 & 0.257 &  5 & 0.129 & 6 &  0.154 & 12 & 0.309 & 20 & 0.515 & 198 & 5.094 & 614 & 15.796 &  1224 & 31.490 & 1791 & 46.077     \\ \hline   
\end{tabular}}
\end{table*}
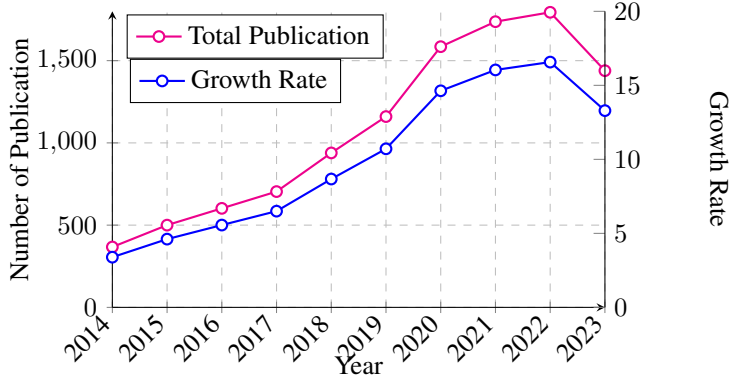
\begin{figure*}
\begin{tikzpicture}
\begin{axis}
[width=0.49\textwidth,
height=5.5cm,
axis lines=left,
xlabel={Year},
xticklabel style = {rotate=45},
symbolic x coords ={2014,2015,2016,2017,2018,2019,2020,2021,2022,2023},
x tick label style={rotate=0,anchor=east},
xtick = data,ylabel={Number of Publication},
ymin = 0, ymax = 1800,
legend style={at={(0.29,0.84)},
anchor=south ,legend columns=2},
ymajorgrids=true,
xmajorgrids=true,
grid style=dashed,thin]
\addplot+ [color=magenta,mark options={fill=white},draw=magenta, thick]
coordinates
{(2014, 367) (2015, 500) (2016,602) (2017,704) (2018,939) (2019,1160) (2020,1585) (2021,1737) (2022,1794) (2023,1439)};
\addlegendentry{Total Publication}
\end{axis}
\pgfplotsset{every axis y label/.append style={rotate=180,yshift=9.25cm}}
\begin{axis}[width=0.49\textwidth,
height=5.5cm,
xmin=2014, xmax=2023,
ymin = 0, ymax = 20,
xtick ={2014,2015,2016,2017,2018,2019,2020,2021,2022,2023},
xtick = data,
hide x axis,
axis y line*=right,
nodes near coords align={vertical},
ylabel={Growth Rate},
legend style={at={(0.25,0.69)},
anchor=south,legend columns=2}]
\addplot+ [color=blue,mark options={fill=white},draw=blue, thick] coordinates {(2014, 3.39) (2015, 4.61) (2016,5.56) (2017,6.50) (2018,8.67) (2019,10.71) (2020,14.63) (2021,16.04) (2022,16.57) (2023,13.29)};
\addlegendentry{Growth Rate}
\end{axis}
\end{tikzpicture}
\caption{Overall Publication Growth}
\end{figure*}
\subsubsection{Domain-wise Temporal Distribution of Publications}
A quantitative analysis approach is employed to outline the fundamental structure of research advancements in each domain.
A total of 12638, documents were retrieved from the WoS database, covering the period from 2014 to 2023. Additionally, all gathered records are categorized based on the cloud computing domains using the WoS subquery. Table III displays the annual growth of each domain, showing the total number of publications (TP) and their percentage share (TP\%) for each respective year \cite{101,102,103,104,105,106,107,108,109,110}. According to the data from the table, the highest number of records are from FL (4133), followed by MEC (3540) followed by FC (2089). In contrast, SCS (1366) reports a minimum number of publications, followed by WS (1407). The analysis reveals that FL is the most extensively researched domain among researchers, while minimal effort is directed towards the SCS domain. Fig. 5 illustrates the overall publication growth across the CC domains graphically.
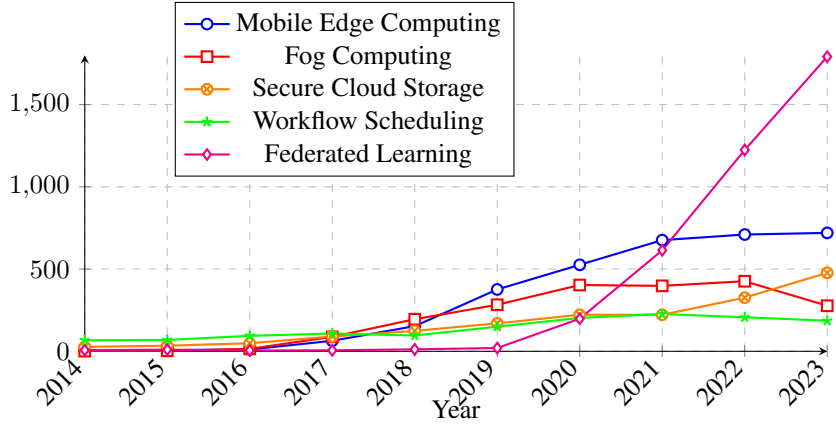
\begin{figure*}
\begin{tikzpicture}
\begin{axis}
[width=0.69\textwidth,
height=5.5cm,
axis lines=left,
xlabel={Year},
xticklabel style = {rotate=45},
symbolic x coords ={2014,2015,2016,2017,2018,2019,2020,2021,2022,2023},
x tick label style={rotate=0,anchor=east},
xtick = data,
ymin = 0, ymax = 1800,
legend style={at={(0.35,0.59)},
anchor=south ,legend columns=1},
ymajorgrids=true,
xmajorgrids=true,
grid style=dashed,thin]
\addplot+ [color=blue,mark options={fill=white},draw=blue, thick]
coordinates
{(2014, 4) (2015, 7) (2016,11) (2017,64) (2018,153) (2019,376) (2020,526) (2021,676) (2022,710) (2023,720)};
\addlegendentry{Mobile Edge Computing}
\addplot+ [color=red,mark options={fill=white},draw=red, thick]
coordinates
{(2014, 1) (2015, 3) (2016,15) (2017,88) (2018,195) (2019,283) (2020,403) (2021,398) (2022,426) (2023,277)};
\addlegendentry{Fog Computing}
\addplot+ [color=orange,mark options={fill=white},draw=orange, thick]
coordinates
{(2014, 27) (2015, 34) (2016,49) (2017,90) (2018,124) (2019,170) (2020,222) (2021,222) (2022,326) (2023,477)};
\addlegendentry{Secure Cloud Storage}
\addplot+ [color=green,mark options={fill=white},draw=green, thick]
coordinates
{(2014, 67) (2015, 69) (2016,94) (2017,108) (2018,96) (2019,150) (2020,203) (2021,227) (2022,207) (2023,186)};
\addlegendentry{Workflow Scheduling}
\addplot+ [color=magenta,mark options={fill=white},draw=magenta, thick]
coordinates
{(2014, 7) (2015, 10) (2016,5) (2017,6) (2018,12) (2019,20) (2020,198) (2021,614) (2022,1224) (2023,1791)};
\addlegendentry{Federated Learning}
\end{axis}
\end{tikzpicture}
\caption{Domainwise overall publication growth in CC}
\end{figure*}
\subsubsection{Geographical Distribution of Publications}
 The geographical distribution analysis or country collaboration analysis reflects both the quality of literature produced and the cooperative relationships between countries in a specific field. This study showcases the country collaboration analysis using the CiteSpace visualization tool across the CC literature. Fig.  4 displays the country collaboration network diagram, comprising 127 nodes and 1211 links. In this diagram, each node represents the number of publications produced in collaboration with other countries, while the links between the two countries indicate the degree of cooperation between them \cite{111,112,113,114,115,116,117,118,119,120,121}. A larger node size suggests a greater publication count for the specified country, whereas a higher number of links indicates stronger inter-country cooperation. The circle surrounding a node, representing the centrality value, illustrates the quality of publications. 
Table IV displays the top 10 countries in the CC literature, listing their rank, frequency of publications, number of neighbor nodes/number of links, and centrality. The analysis indicates that England demonstrates the greatest degree of cooperation among countries, with a maximum of 73 links between nodes in the country collaboration network. Peoples R China, USA, India, South Korea, and Australia are the five countries with the highest intercountry cooperation, followed by Canada. Furthermore, the analysis indicates that Peoples R China is the leading contributor to the CC literature, with 5656 publications. USA and India are the two most significant contributors, followed by Peoples R China; then, South Korea, and Australia hold fourth and fifth positions, respectively. 
From the context of centrality, England (0.16), Saudi Arabia (0.14), and the USA (0.12) are the top three countries, indicating the highest quality publications. Some countries, such as England (0.16) and Saudi Arabia (0.14), possess lower publications count but higher centrality, indicating that the CC literature published in these countries is less but of higher quality. In contrast, the publications count of Peoples R China (5616) is remarkable, but the low centrality (0.03) indicates the requirement of the researcher’s efforts on quality enhancement.
\begin{figure*}[htbp]
  \centering
\includegraphics[width=1.2\textwidth]{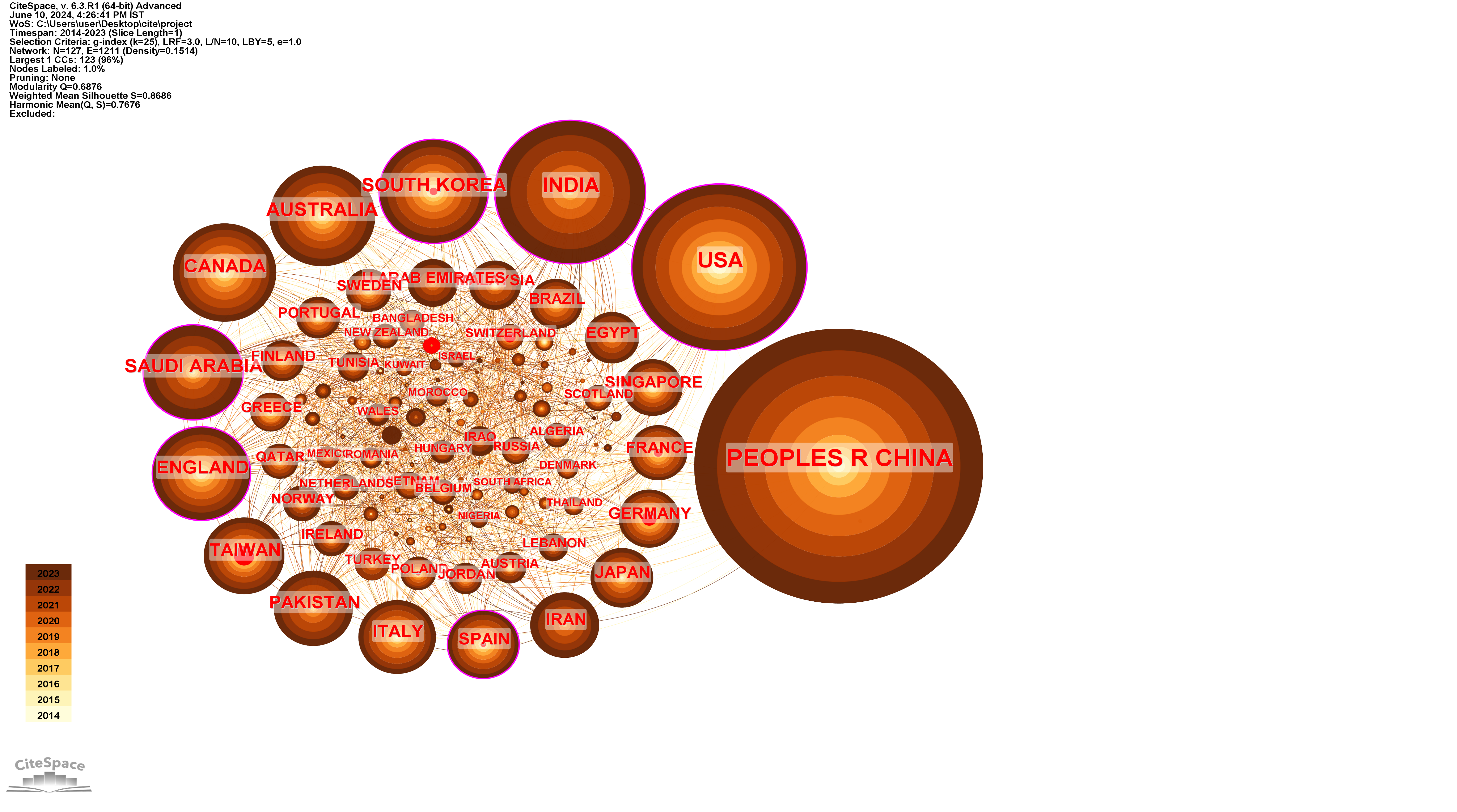}
\caption{ Overall country collaboration network}
\label{fig}
\end{figure*}
\begin{table} [!htbp]
    \centering
    \caption{Country Co-citation Analysis }
    \resizebox{\columnwidth}{!}{
    \begin{tabular}{ |c|c|l|c|c| } 
    \hline
    Rank & Frequency & Country & Link &  Centality\\ [0.5 ex]
    \hline 
    1 & 5616 & PEOPLES R CHINA & 41 &  0.03 \\ \hline
    2 & 2060 & USA & 63  & 0.12 \\ \hline
    3 &  1548 & INDIA  & 49  & 0.10  \\ \hline
    4  & 821  & SOUTH KOREA & 44 & 0.11 \\ \hline
    5  & 792 &  AUSTRALIA & 52 & 0.08\\ \hline
    6  & 763 & CANADA & 44 & 0.04\\ \hline
    7  & 680 & SAUDI ARABIA & 61 & 0.14\\ \hline
    8  & 650 & ENGLAND & 73 & 0.16\\ \hline
    9  & 470 &  TAIWAN & 32 & 0.02\\ \hline
    10  & 460 & PAKISTAN & 45& 0.09\\ \hline  
\end{tabular}}
\label {table:1}   
 \end{table}
\subsection{Co-Citation Analysis}

\subsubsection{Documentation co-citation Analysis}
Documents serve as repositories of critical knowledge. When two or more documents are cited simultaneously by one or more subsequent documents, they share a co-citation relationship \cite{122,123,124,125,126,127,128,129,130,131,132}. 
Document co-citation analysis allows us to objectively investigate the underlying knowledge base of a research field, as well as its knowledge structure and evolutionary path. In the document co-citation network, the nodes representing cited documents are labeled with the first author and the year of publication. The links indicate the co-citation relationships between these documents. The node size represents the importance of a document, while the distance between nodes signifies the frequency of their co-citations. A smaller distance between nodes indicates a higher co-citation frequency and greater similarity in research topics. The network in Fig. 7 comprises 1422 nodes and 6929 links, and the network density is 0.0069. Table IV lists the top 10 cited documents; from the citation count, Shi WS, Abbas N, and Mach P are the top three cited articles. These articles have been widely recognized by peers and have high value in cloud computing. 
\begin{figure*}[!htbp]
  \centering
\includegraphics[width=1.29\textwidth]{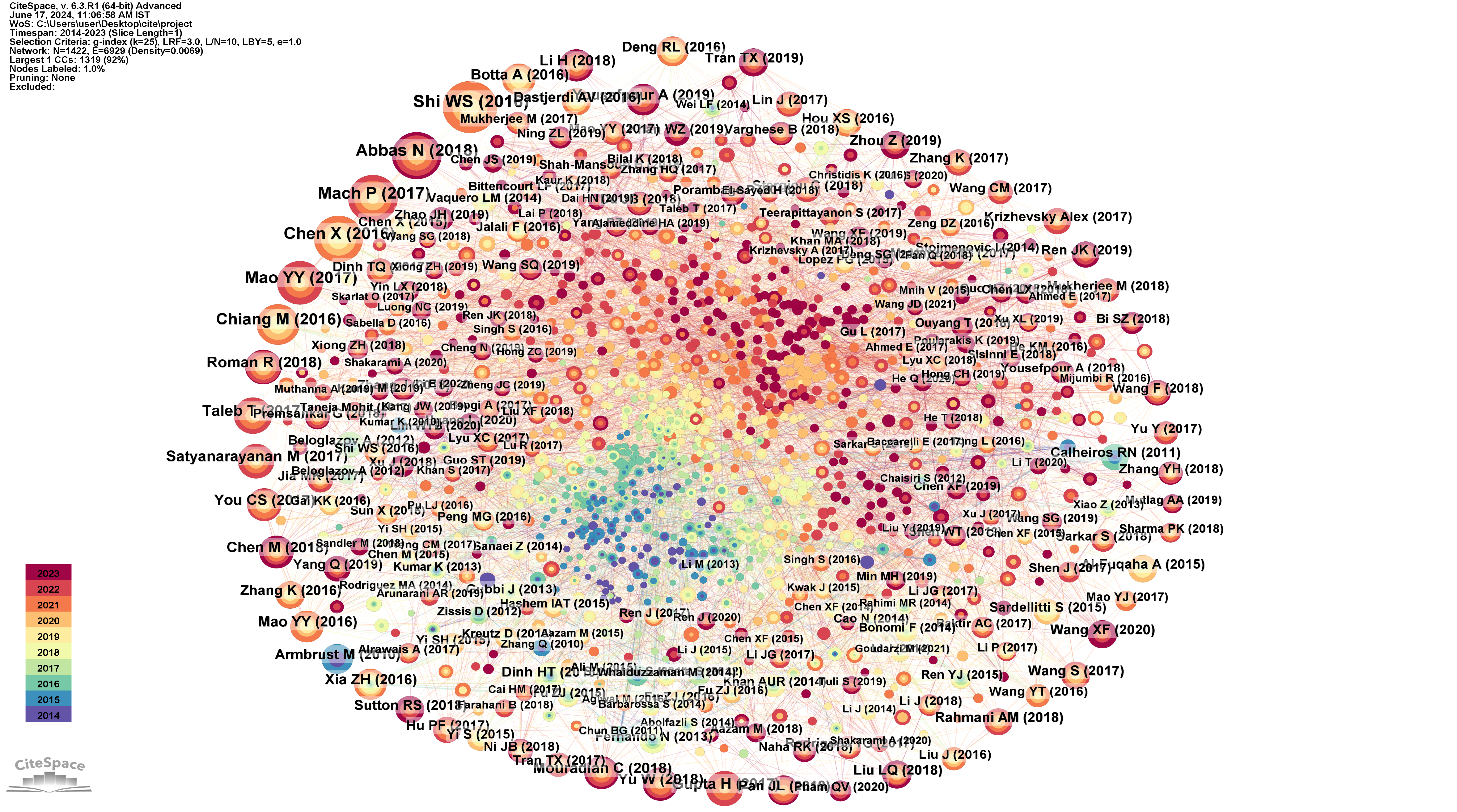}
\caption{Document co-citation network}
\label{fig}
\end{figure*}
\begin{table}[!htbp]
\centering
\caption{Top 10 co-cited documents in Cloud Computing}
\resizebox{0.49\textwidth}{!}{
\begin{tabular}{p{0.35cm}p{0.35cm}cp{1.45cm}lp{2.35cm}c}
\hline
Rank & Count & Degree & Author & Document & Journal & Year \\
\hline
1& 314 & 23 & Shi WS  & 10.1109/JIOT.2016.2579198   & IEEE INTERNET THINGS & 2016\\
2& 276 & 34 & Abbas N  & 10.1109/JIOT.2017.2750180   & IEEE INTERNET THINGS & 2018\\
3& 265 & 54 & Mach P   & 10.1109/COMST.2017.2682318  & IEEE COMMUN SURV TUT & 2017\\
4& 257 & 70 & Chen X   & 10.1109/TNET.2015.2487344  & IEEE ACM T NETWORK  & 2016\\
5& 223 & 37 & Mao YY   & 10.1109/COMST.2017.2745201	   & IEEE COMMUN SURV TUT  & 2017\\
6& 199 & 40 & Chiang M & 10.1109/JIOT.2016.2584538  & IEEE INTERNET THINGS  & 2016\\
7& 164 & 27 & Taleb T  & 10.1109/COMST.2017.2705720	 & IEEE COMMUN SURV TUT & 2017\\
8 & 144 & 24 & Gupta H  & 10.1002/spe.2509 & SOFTWARE PRACT EXPER& 2018\\
9 & 144 & 22 & Roman R  & 10.1016/j.future.2016.11.009 & FUTURE GENER COMP SY  & 2017\\
10& 142 & 11 & Satyanarayanan M & 10.1109/MC.2017.9 & COMPUTER & 2017\\
\hline
\end{tabular}}
\end{table}
\subsubsection{Author co-citation Analysis}
Author co-citation analysis (ACA) is a technique employed to identify and visualize the intellectual structure within a specific academic discipline \cite{133,134,135,136,137,138,139,140,141,142,143}. ACA involves counting how often the works of one author are cited together with the works of another author in the references of citing documents. This analysis helps to reveal relationships among authors whose publications are frequently cited together in scholarly articles. 
Fig. 8 displays the author's co-citation network contributing to the cloud computing research domain, comprising 649 nodes and 1194 co-citation links. In the author co-citation network, node size represents each author's co-citation frequency, while the links indicate indirect collaborative relationships based on their co-citation frequency. Table V lists the top 10 cited authors and those with the highest centrality based on the citation frequency of articles related to the cloud computing domain. As shown in Table V, the most cited author is Buyya, Rajkumar (83), and the other top 9 cited authors are Choo, Kim-Kwang Raymond (71), Liu, Ximeng  (67), Guizani, Mohsen (66), Ma, Jianfeng (58), Li, Keqin (51), Wang, Shangguang (44), Yang, Yang  (42), Jin, Hai (41), and Yang, Laurence T  (41). In addition, the three cited authors with the highest centrality are Choo, Kim-Kwang Raymond  (0.17), Guizani, Mohsen  (0.1), and Buyya, Rajkumar (0.08), indicating that these authors are key nodes in the field of cloud computing research and play a role in high linkage.
\begin{figure*}[htbp]
  \centering
\includegraphics[width=1\textwidth]{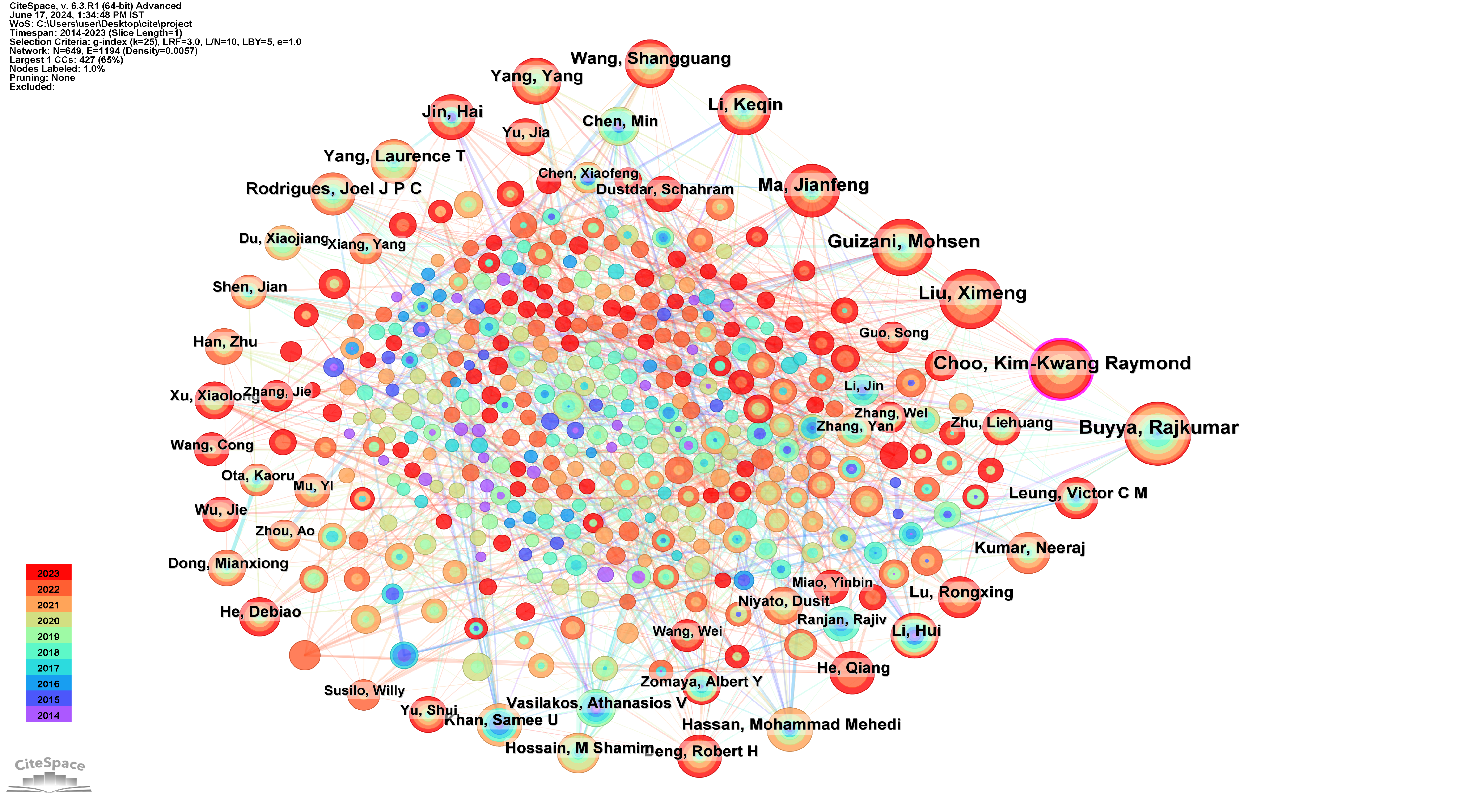}
\caption{Author co-citation network}
\label{fig}
\end{figure*}
\begin{table}[htbp]
\centering
\caption{High-yield authors in the field of Cloud computing}
\begin{tabular}{llccc}
\hline
Rank & Author & Frequency & Year & Centrality \\
\hline
1& Buyya, Rajkumar& 83&  2014& 0.08\\ 
2& Choo, Kim-Kwang Raymond & 71 &2015 & 0.17\\ 
3& Liu, Ximeng & 67 & 2018 & 0.07\\ 
4& Guizani, Mohsen & 66 & 2014  & 0.1\\ 
5& Ma, Jianfeng & 58 & 2015 & 0.03\\ 
6& Li, Keqin & 51 &2014 & 0.04\\ 
7& Wang, Shangguang & 44 &2016 & 0.04\\ 
8& Yang, Yang & 42 & 2016 & 0.03\\ 
9 & Jin, Hai & 41  &2014 & 0.02\\ 
10& Yang, Laurence T & 41   &2014 & 0.04\\ 
\hline
\end{tabular}
\end{table}
\subsubsection{Journal co-citation Analysis}
Analyzing the citation patterns of journals in scientometrics helps researchers identify the most influential journals in a specific field, providing them with a platform for impactful scientific writing \cite{145,146,147,148,149,150,151,152,153,154,155}. This analysis determines the most prestigious journals by considering the total citation count.
The network in Fig. 8 contains 1743 nodes and 13275 links. The selection criteria are the top 50 per slice, and the network density of the journal co-citation network is 0.0087. As analyzed above, the generated journal co-citation network is comprehensive, and the co-citation relationship of journals is strong. Table VI shows the list of highly cited journals with the citation count, degree, and Impact Factor. The analysis depicts that IEEE ACCESS has the highest number of citations (5154) in the cloud computing domain. Hence, IEEE ACCESS is the most followed journal among researchers with the maximum publications in the domain of CC. IEEE COMMUNICATIONS SURVEYS AND TUTORIALS, IEEE TRANSACTIONS ON INDUSTRIAL INFORMATICS, and IEEE COMMUNICATIONS MAGAZINE are the journals with the highest impact factor, i.e., 35.6, 12.3, and 11.2, respectively. As discussed above, the journal co-citation analysis provides the distribution of the sources of critical knowledge in cloud computing, which could help us identify the journals that are cited and determine the core journals and the connections between them. 
\begin{figure*}[!htbp]
  \centering
\includegraphics[width=\textwidth]{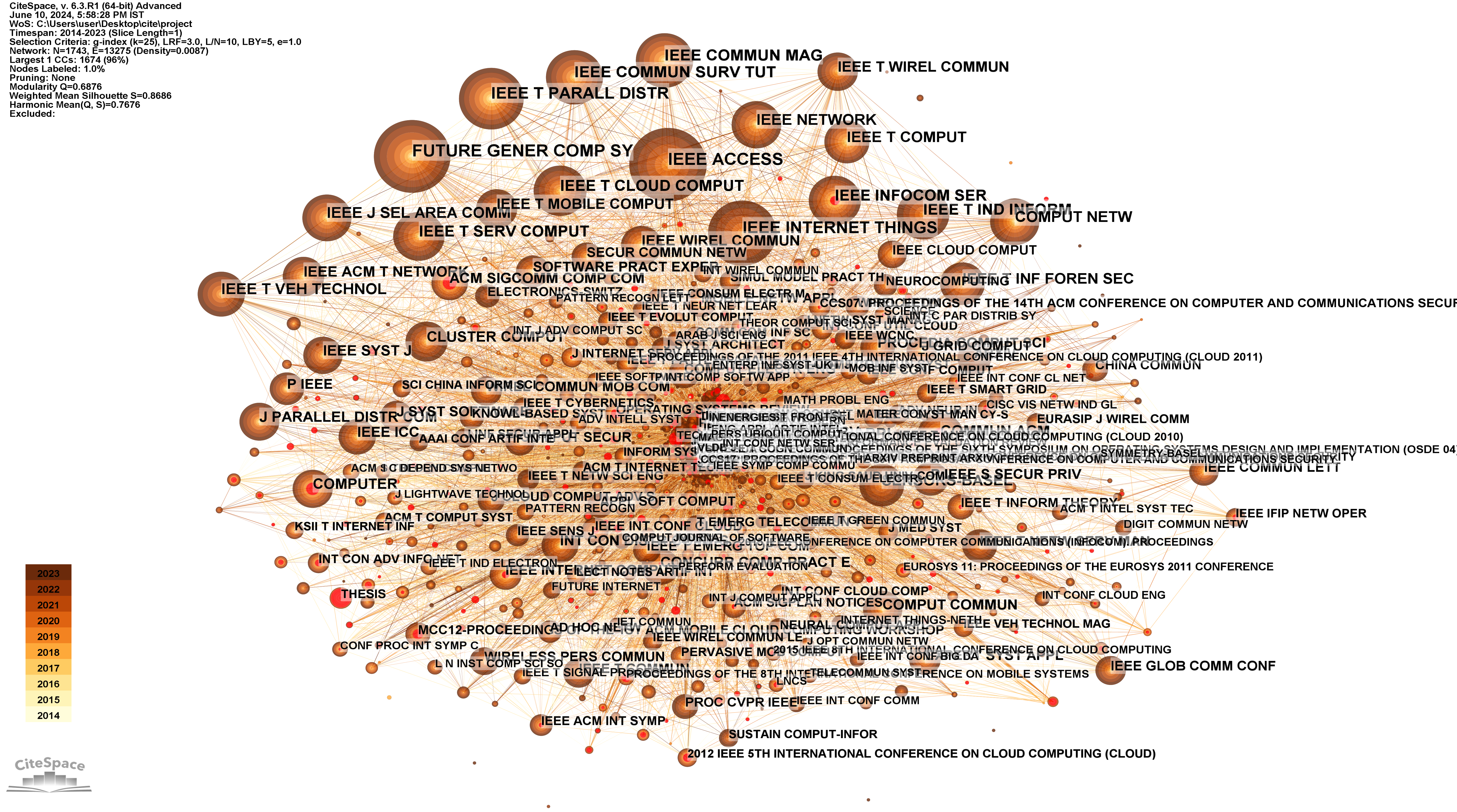}
\caption{Journal Co-citation Network}
\label{fig}
\end{figure*}
\begin{table}[htbp]
\centering
\caption{ Top 10 and high centrality cited journals of Cloud computing}
\begin{tabular}{p{0.35cm}cp{0.5cm}p{3.75cm}c}
\hline
Rank & Frequency & Degree & Journal & IF \\
\hline
1& 5154 & 127 & IEEE ACCESS  & 3.9 \\ 
2 & 5046 & 123 & FUTURE GENERATION COMPUTER SYSTEM & 7.5 \\
3 & 4095 & 123 & IEEE INTERNET THINGS  & 10.6  \\
4 & 3645 & 80 & IEEE TRANSACTIONS ON PARALLEL DISTRIBUTED SYSTEMS & 5.3 \\
5 & 2879 & 85 & IEEE COMMUNICATIONS MAGAZINE & 11.2  \\
6 & 2855 & 96 & IEEE COMMUNICATIONS SURVEYS AND TUTORIALS & 35.6  \\
7 & 2683 & 94 & JOURNAL OF NETWORK COMPUTER APPLICATIONS & 8.7  \\
8 & 2427 & 71 & IEEE TRANSACTIONS ON CLOUD COMPUTING & 6.5 \\
9 & 2417 & 98 & IEEE TRANSACTIONS ON INDUSTRIAL INFORMATICS & 12.3  \\
10 & 2336 & 67 & IEEE INFOCOM SER & 4.68  \\
\hline
\end{tabular}
\end{table}
\subsection{Co-Word Analysis}
 Keywords serve as effective tools for abstractly representing and categorizing the content of a scientific article. From a meta-perspective, they form the foundation for analyzing the key topics and aspects that define a particular research area. New popular topics can be identified rapidly by observing the frequency of keywords within a defined time frame. Analyzing the co-occurrence of keywords also helps in identifying closely related topics or aspects.
 
\subsubsection{Keyword co-citation Analysis}
Table VII presents the ranking of keywords based on their high frequency.  The results indicate that recent research activities are mainly focused on the utilization of scalable cloud resources such as resource allocation, optimization, resource management, etc. Popular links to other fields of research are also revealed, for instance with the keywords internet of things, edge computing, and fog computing. By analyzing frequent keywords per year, the emergence and growing popularity of specific topics can be documented. For example, the growing importance of cloud computing to efficiently process large and complex masses of structured and unstructured data is depicted by keywords shown in Table VII. From the obtained result, we recognize the emergence of new research topics in the field of cloud computing, such as Big Data and the Internet of Things. Consequently, keyword analysis can be used as a tool for identifying current research trends. The top 25 keywords with citation bursts are shown in
 Figure 11. Considering the burst strength and its corresponding red lines, the evolution of research frontiers is shown.
\begin{figure*}[htbp]
  \centering
\includegraphics[width=1\textwidth]{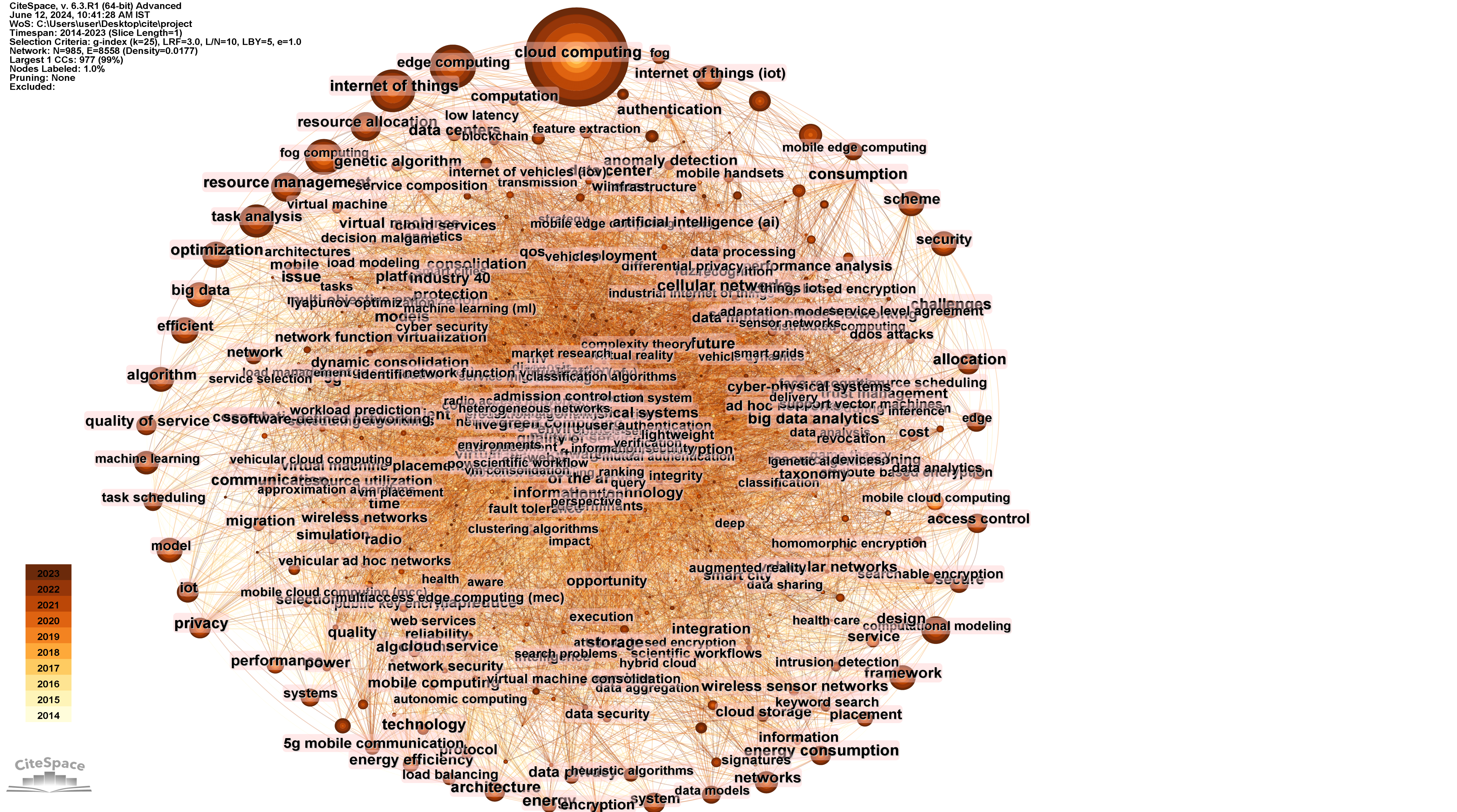}
\caption{Keyword co-citation Analysis}
\label{fig}
\end{figure*}
\begin{table}[htbp]
\centering
\caption{Keywords Frequency Distribution}
\begin{tabular}{llc}
\hline
No. & Keyword & Frequency  \\
\hline
1 & cloud computing & 5713\\ 
2 & edge computing & 1329 \\ 
3 & internet of things & 1261\\ 
4 & resource allocation & 792\\ 
5 & fog computing & 791\\ 
6 & task analysis & 668 \\ 
7 & resource management & 660 \\ 
8 & algorithm & 615\\ 
9 & optimization & 592\\ 
10 & big data & 591\\
11 & security & 531\\
12 & efficient & 519\\
13 & scheme & 498\\
14 & model & 480\\
15 & machine learning & 389\\
16 &privacy & 368 \\
17 & access control & 345\\
18 & energy consumption & 339\\
19 & secure & 330\\
20 & quality of service & 321\\
\hline
\end{tabular}
\end{table}%
\begin{figure*}[htbp]
  \centering
\includegraphics[width=0.9\textwidth]{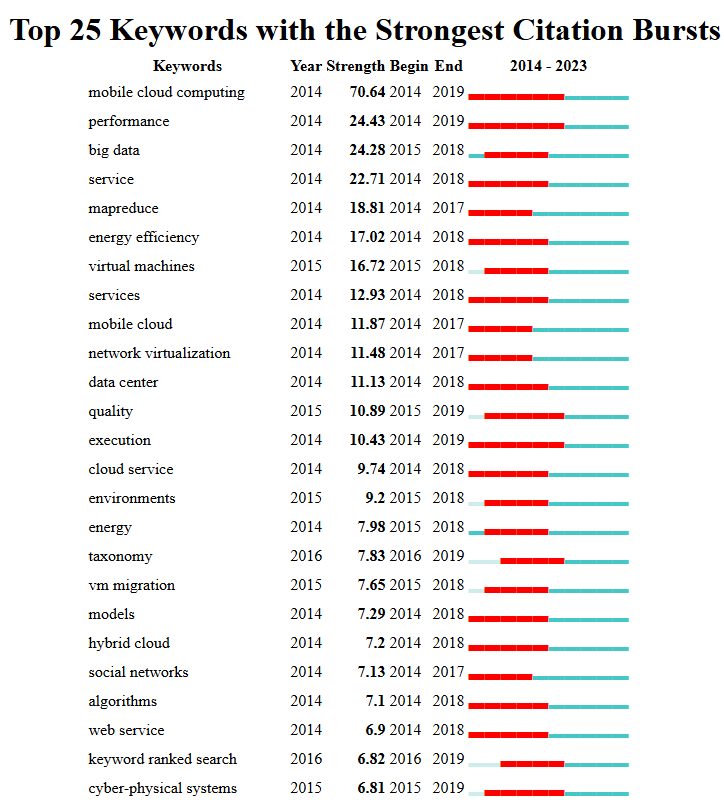}
\caption{Top 25 keywords with citation bursts.}
\label{fig}
\end{figure*}
\section{Conclusion, And Future Directions}
Cloud computing attracts significant interdisciplinary attention and is a rapidly developing field of research. In this paper, we conduct a scientometric analysis to comprehensively investigate the development and current state of cloud computing-related publications using a large bibliographic dataset provided by the Web of Science database. The results of this study reveal that the focus of research activities is predominantly influenced by fundamental and highly recognized scientists and publications.
From 2014 to 2023, the number of research papers on cloud computing increased significantly. Regarding contributions to cloud computing research, from an author -perspective, the high-yield authors are Buyya Rajkumar, Choo Kim-Kwang Raymond, and Liu Ximeng. Thus, Choo Kim-Kwang Raymond  is both a high-yield
and a highly cited author. From the country perspective, PEOPLES R CHINA, the USA, INDIA, SOUTH KOREA, and AUSTRALIA  have high-yielding institutions in the field. The PEOPLES R CHINA has a high number of publications in the country cooperation network and England and SAUDI ARABIA have high centrality. Regarding the core journals, high-yielding journals include IEEE ACCESS, FUTURE GENERATION COMPUTER SYSTEM, IEEE INTERNET THINGS, and IEEE TRANSACTIONS ON PARALLEL DISTRIBUTED SYSTEMS. At the same time, highly cited journal is the IEEE ACCESS; most of these have high quality and are sources of knowledge for the study of cloud computing. According to the co-word analysis of high-frequency 
keywords in cloud computing literature, cloud computing, edge computing, internet of things, resource allocation, fog computing,  task analysis, and resource management have higher frequency counts, more than 500 times.  
\par
This research can be further extended by incorporating institution co-occurrence analysis. Scholars can utilize additional data visualization tools and databases to gain valuable insights, uncover patterns, and make informed decisions for future research and exploration in the field of CC. Additionally, examining the research contributions of subdomains can enhance the depth of the study.
\bibliographystyle{IEEEtran}
\bibliography{references}

\begin{thebibliography}{100}
\providecommand{\url}[1]{#1}
\csname url@samestyle\endcsname
\providecommand{\newblock}{\relax}
\providecommand{\bibinfo}[2]{#2}
\providecommand{\BIBentrySTDinterwordspacing}{\spaceskip=0pt\relax}
\providecommand{\BIBentryALTinterwordstretchfactor}{4}
\providecommand{\BIBentryALTinterwordspacing}{\spaceskip=\fontdimen2\font plus
\BIBentryALTinterwordstretchfactor\fontdimen3\font minus
  \fontdimen4\font\relax}
\providecommand{\BIBforeignlanguage}[2]{{%
\expandafter\ifx\csname l@#1\endcsname\relax
\typeout{** WARNING: IEEEtran.bst: No hyphenation pattern has been}%
\typeout{** loaded for the language `#1'. Using the pattern for}%
\typeout{** the default language instead.}%
\else
\language=\csname l@#1\endcsname
\fi
#2}}
\providecommand{\BIBdecl}{\relax}
\BIBdecl

\bibitem{1}
D.~Saxena and A.~K. Singh, ``Concepts, taxonomic review, and emerging trends in
  computational intelligence for green cloud systems,'' \emph{Computer Science
  Review}, vol.~60, p. 100894, 2026.

\bibitem{2}
P.~Rani, A.~K. Singh, A.~Parashar, and D.~Saxena, ``Quantum-based multifaceted
  cybersecurity model for smart grid data communications,'' \emph{IEEE
  Transactions on Automation Science and Engineering}, vol.~23, pp. 7778--7793,
  2026.

\bibitem{3}
D.~Saxena and A.~K. Singh, ``Multifactor trust-driven secure communication
  model for cloud-based digital twins,'' \emph{IEEE Transactions on Industrial
  Informatics}, pp. 1--10, 2026.

\bibitem{4}
K.~Gupta, D.~Saxena, A.~K. Singh, and C.-N. Lee, ``Neighbor-embedded graph
  neural network-based crowd delivery traffic management in smart city,''
  \emph{IEEE Transactions on Emerging Topics in Computational Intelligence},
  pp. 1--13, 2026.

\bibitem{5}
D.~Saxena, A.~Rajput, and A.~Kumar~Singh, ``Unleashing the power of artificial
  intelligence for exploring unrevealed and unexplored natural resources,''
  \emph{IT Professional}, vol.~28, no.~1, pp. 35--43, 2026.

\bibitem{6}
R.~N. Wedamuni~Arachchige, D.~Saxena, and A.~K. Singh, ``An adaptive cyber
  threat intelligence model to counter evolving security attacks in industrial
  communication networks,'' \emph{Neural Computing and Applications}, vol.~38,
  no.~2, p.~15, 2026.

\bibitem{7}
V.~Mishra, D.~Saxena, and A.~K. Singh, ``Digital health challenges and
  opportunities for health-care it professionals,'' \emph{IT Professional},
  vol.~27, no.~6, pp. 65--70, 2025.

\bibitem{8}
P.~Rani, A.~K. Singh, A.~Parashar, and D.~Saxena, ``Quantum fourier
  transformation and clifford gate-driven secure communication model for smart
  grid environments,'' \emph{IEEE Transactions on Dependable and Secure
  Computing}, vol.~23, no.~2, pp. 3681--3695, 2026.

\bibitem{9}
D.~Saxena and A.~K. Singh, ``A meta-unified global cyber threat intelligence
  model for industrial cross-cloud networks,'' \emph{IEEE Transactions on
  Information Forensics and Security}, vol.~20, pp. 12\,317--12\,327, 2025.

\bibitem{10}
D.~Saxena, H.~M. Gaur, A.~K. Singh, and A.~Mohan, ``Quantum blackhole
  learning-optimized hadamard neural network model for dynamic resource
  reservation in industry clouds,'' \emph{IEEE Transactions on Systems, Man,
  and Cybernetics: Systems}, vol.~56, no.~1, pp. 134--147, 2026.

\bibitem{11}
L.-W. Li, C.-N. Lee, K.~Gupta, H.-F. Yang, and A.~Kumar~Singh, ``Syntax element
  encryption for h.265/hevc using chaotic map-based coefficient scrambling
  scheme,'' \emph{IEEE Transactions on Circuits and Systems for Video
  Technology}, vol.~36, no.~4, pp. 5655--5670, 2026.

\bibitem{12}
N.~Singh, K.~Gupta, A.~K. Singh, P.~Nallagownden, and I.~Elamvazuthi,
  ``Reinforcement learning based multi-agent system for smart microgrid,''
  \emph{Journal of Network and Computer Applications}, p. 104339, 2025.

\bibitem{13}
V.~Mishra, D.~Saxena, K.~Gupta, S.~Patni, and A.~K. Singh, ``Sustainability in
  large language model supply chains-insights and recommendations using
  analysis of utility for affecting factors,'' \emph{Scientific Reports},
  vol.~15, no.~1, p. 33524, 2025.

\bibitem{14}
D.~Saxena, A.~K. Singh, and V.~Lindenstruth, ``Quad-caching management model
  for heterogeneous data lake environments,'' \emph{Expert Systems with
  Applications}, p. 129133, 2025.

\bibitem{15}
D.~Saxena and A.~K. Singh, ``An intelligent secure and reliable cloud services
  management model with toffoli gate-embedded quantum adam neural network,''
  \emph{IEEE Transactions on Dependable and Secure Computing}, vol.~22, no.~6,
  pp. 6554--6565, 2025.

\bibitem{16}
A.~K. Singh, D.~Saxena, and V.~Lindenstruth, ``Ree-tm: Reliable and
  energy-efficient traffic management model for diverse cloud workloads,''
  \emph{IEEE Transactions on Cloud Computing}, vol.~13, no.~3, pp. 953--968,
  2025.

\bibitem{17}
J.~Kumar, D.~Saxena, K.~Gupta, S.~Kumar, and A.~K. Singh, ``A comprehensively
  adaptive architectural optimization-ingrained quantum neural network model
  for cloud workloads prediction,'' \emph{IEEE Transactions on Neural Networks
  and Learning Systems}, vol.~36, no.~10, pp. 19\,039--19\,053, 2025.

\bibitem{18}
J.~Kumar, P.~Rani, D.~Saxena, A.~K. Singh, and A.~Makkar, ``Evolutionary
  learning driven load forecasting and demand response management model for
  smart grid,'' \emph{Applied Soft Computing}, vol. 177, p. 113169, 2025.

\bibitem{19}
D.~Saxena, N.~Singh, K.~Gupta, A.~Verma, V.~Mishra, J.~Kumar, I.~Gupta,
  S.~Patni, R.~Gupta, J.~Kumar, and A.~K. Singh, ``An intelligent multi-depot
  vehicle routing and management model for smart cities,'' \emph{IEEE
  Transactions on Intelligent Transportation Systems}, vol.~26, no.~6, pp.
  7740--7754, 2025.

\bibitem{20}
A.~K. Singh, S.~Kumar, and S.~Jain, ``A multi-agent deep reinforcement learning
  approach for optimal resource management in serverless computing,''
  \emph{Cluster Computing}, vol.~28, no.~2, p. 102, 2025.

\bibitem{31}
D.~Saxena, I.~Gupta, J.~Kumar, A.~K. Singh, and X.~Wen, ``A secure and
  multiobjective virtual machine placement framework for cloud data center,''
  \emph{IEEE Systems Journal}, 2021.

\bibitem{32}
N.~K. Sharma and G.~R.~M. Reddy, ``Multi-objective energy efficient virtual
  machines allocation at the cloud data center,'' \emph{IEEE Transactions on
  Services Computing}, vol.~12, no.~1, pp. 158--171, 2016.

\bibitem{33}
J.~Han, W.~Zang, S.~Chen, and M.~Yu, ``Reducing security risks of clouds
  through virtual machine placement,'' in \emph{IFIP Annual Conference on Data
  and Applications Security and Privacy}.\hskip 1em plus 0.5em minus
  0.4em\relax Springer, 2017, pp. 275--292.

\bibitem{34}
A.~K. Singh and J.~Kumar, ``Secure and energy aware load balancing framework
  for cloud data centre networks,'' \emph{Electronics Letters}, vol.~55, no.~9,
  pp. 540--541, 2019.

\bibitem{35}
D.~Saxena and A.~K. Singh, ``A proactive autoscaling and energy-efficient vm
  allocation framework using online multi-resource neural network for cloud
  data center,'' \emph{Neurocomputing}, vol. 426, pp. 248--264, 2021.

\bibitem{36}
------, ``Energy aware resource efficient-(eare) server consolidation framework
  for cloud datacenter,'' in \emph{Advances in communication and computational
  technology}.\hskip 1em plus 0.5em minus 0.4em\relax Springer, 2021, pp.
  1455--1464.

\bibitem{37}
------, ``Communication cost aware resource efficient load balancing (carelb)
  framework for cloud datacenter,'' \emph{Recent Advances in Computer Science
  and Communications (Formerly: Recent Patents on Computer Science)}, vol.~14,
  no.~9, pp. 2920--2933, 2021.

\bibitem{38}
J.~Kumar and A.~K. Singh, ``Workload prediction in cloud using artificial
  neural network and adaptive differential evolution,'' \emph{Future Generation
  Computer Systems}, vol.~81, pp. 41--52, 2018.

\bibitem{39}
J.~Kumar, R.~Goomer, and A.~K. Singh, ``Long short term memory recurrent neural
  network (lstm-rnn) based workload forecasting model for cloud datacenters,''
  \emph{Procedia Computer Science}, vol. 125, pp. 676--682, 2018.

\bibitem{40}
D.~Saxena, I.~Gupta, A.~K. Singh, and C.-N. Lee, ``A fault tolerant elastic
  resource management framework towards high availability of cloud services,''
  \emph{IEEE Transactions on Network and Service Management}, 2022.

\bibitem{21}
J.~Kumar, D.~Saxena, J.~Kumar, A.~Bhadoria, A.~Anjali, and A.~K. Singh, ``A
  hybrid neural network and cooperative pso model for dynamic cloud workloads
  prediction,'' \emph{Computing}, vol. 107, no.~3, p.~72, 2025.

\bibitem{22}
D.~Saxena and A.~K. Singh, ``A self-healing and fault-tolerant cloud-based
  digital twin processing management model,'' \emph{IEEE Transactions on
  Industrial Informatics}, vol.~21, no.~5, pp. 4233--4242, 2025.

\bibitem{23}
S.~Patni, J.~Kumar, D.~Saxena, and A.~Kumar~Singh, ``Classified dynamic
  hierarchical load balancer for cloud data centers,'' \emph{SN Computer
  Science}, vol.~6, no.~3, p. 218, 2025.

\bibitem{24}
S.~Patni, D.~Saxena, and A.~K. Singh, ``Data security and leakage detection
  models,'' in \emph{Resource Management in Cloud Computing: Concepts and
  Implementation}.\hskip 1em plus 0.5em minus 0.4em\relax Springer, 2025, pp.
  159--189.

\bibitem{25}
------, ``Secure and energy-efficient cloud traffic management schemes,'' in
  \emph{Resource Management in Cloud Computing: Concepts and
  Implementation}.\hskip 1em plus 0.5em minus 0.4em\relax Springer, 2025, pp.
  135--157.

\bibitem{26}
S.~K. Sood, Y.~S. Lamba, and A.~K. Singh, ``Exploring the shift from 3-d
  printing to 4-d printing: A review,'' \emph{IEEE Sensors Journal}, vol.~25,
  no.~6, pp. 9224--9232, 2025.

\bibitem{27}
D.~Saxena and A.~K. Singh, ``Multi-vm-workloads learning-based forecasting
  model for resilient server management in industry clouds,'' \emph{Procedia
  Computer Science}, vol. 260, pp. 923--929, 2025.

\bibitem{28}
S.~R. Swain, A.~Parashar, A.~K. Singh, and C.~N. Lee, ``An intelligent virtual
  machine allocation optimization model for energy-efficient and reliable cloud
  environment: Sr swain et al.'' \emph{The Journal of Supercomputing}, vol.~81,
  no.~1, p. 237, 2025.

\bibitem{29}
A.~K. Singh, S.~R. Swain, D.~Saxena, and C.-N. Lee, ``A bio-inspired virtual
  machine placement toward sustainable cloud resource management,'' \emph{IEEE
  Systems Journal}, 2023.

\bibitem{30}
D.~Saxena, A.~K. Singh, and R.~Buyya, ``Op-mlb: an online vm prediction-based
  multi-objective load balancing framework for resource management at cloud
  data center,'' \emph{IEEE Transactions on Cloud Computing}, vol.~10, no.~4,
  pp. 2804--2816, 2021.

\bibitem{sivakumaren2012growth}
K.~Sivakumaren, S.~Swaminathan, and G.~Karthikeyan, ``Growth and development of
  publication on cloud computing: A scientometric study,'' \emph{International
  Journal of Information Library and Society}, vol.~1, no.~1, p.~37, 2012.

\bibitem{bai2011scientometric}
Q.~Bai and W.-h. Dong, ``Scientometric analysis on the papers of cloud
  computing,'' \emph{Sci-Tech Inform. Develop. Econ.}, vol.~5, no.~1, pp. 6--8,
  2011.

\bibitem{41}
I.~Gupta, D.~Saxena, A.~K. Singh, and C.-N. Lee, ``A multiple controlled
  toffoli driven adaptive quantum neural network model for dynamic workload
  prediction in cloud environments,'' \emph{IEEE Transactions on Pattern
  Analysis and Machine Intelligence}, pp. 1--16, 2024.

\bibitem{42}
S.~R. Swain, D.~Saxena, J.~Kumar, A.~K. Singh, and C.-N. Lee, ``An intelligent
  straggler traffic management framework for sustainable cloud environments,''
  \emph{IEEE Transactions on Sustainable Computing}, pp. 1--13, 2024.

\bibitem{43}
D.~Saxena and A.~K. Singh, ``A high up-time and security centered resource
  provisioning model towards sustainable cloud service management,'' \emph{IEEE
  Transactions on Green Communications and Networking}, pp. 1--1, 2024.

\bibitem{44}
D.~Saxena, J.~Kumar, A.~K. Singh, and S.~Schmid, ``Performance analysis of
  machine learning centered workload prediction models for cloud,'' \emph{IEEE
  Transactions on Parallel and Distributed Systems}, vol.~34, no.~4, pp.
  1313--1330, 2023.

\bibitem{45}
A.~K. Singh, S.~R. Swain, D.~Saxena, and C.-N. Lee, ``A bio-inspired virtual
  machine placement toward sustainable cloud resource management,'' \emph{IEEE
  Systems Journal}, vol.~17, no.~3, pp. 3894--3905, 2023.

\bibitem{46}
J.~Kumar, D.~Saxena, A.~K. Singh, and A.~V. Vasilakos, ``A quantum
  controlled-not neural network-based load forecast and management model for
  smart grid,'' \emph{IEEE Systems Journal}, vol.~17, no.~4, pp. 5714--5725,
  2023.

\bibitem{47}
D.~Saxena and A.~K. Singh, ``A comprehensive survey on sustainable resource
  management in cloud computing environments,'' \emph{Authorea Preprints},
  2024.

\bibitem{48}
S.~Chhabra and A.~K. Singh, ``Secure and energy efficient dynamic hierarchical
  load balancing framework for cloud data centers,'' \emph{Multimedia Tools and
  Applications}, vol.~82, no.~19, pp. 29\,843--29\,856, 2023.

\bibitem{49}
S.~R. Swain, A.~Parashar, A.~K. Singh, and C.~N. Lee, ``Efficient straggler
  task management in cloud environment using stochastic gradient descent with
  momentum learning-driven neural networks,'' \emph{Cluster Computing}, pp.
  1--13, 2023.

\bibitem{50}
A.~K. Singh, D.~Saxena, J.~Kumar, and V.~Gupta, ``A quantum approach towards
  the adaptive prediction of cloud workloads,'' \emph{IEEE Transactions on
  Parallel and Distributed Systems}, vol.~32, no.~12, pp. 2893--2905, 2021.

\bibitem{wang2013research}
T.~Wang and G.-b. Huang, ``Research progress of cloud security from 2008 to
  2011 in china,'' \emph{Inform. Sci.}, 2013.

\bibitem{heilig2014scientometric}
L.~Heilig and S.~Vo{\ss}, ``A scientometric analysis of cloud computing
  literature,'' \emph{IEEE Transactions on Cloud Computing}, vol.~2, no.~3, pp.
  266--278, 2014.

\bibitem{51}
J.~Kumar and A.~K. Singh, ``Dynamic resource scaling in cloud using neural
  network and black hole algorithm,'' in \emph{2016 Fifth International
  Conference on Eco-friendly Computing and Communication Systems
  (ICECCS)}.\hskip 1em plus 0.5em minus 0.4em\relax IEEE, 2016, pp. 63--67.

\bibitem{52}
------, ``Cloud datacenter workload estimation using error preventive time
  series forecasting models,'' \emph{Cluster Computing}, vol.~23, no.~2, pp.
  1363--1379, 2020.

\bibitem{53}
I.~Gupta, R.~Gupta, A.~K. Singh, and R.~Buyya, ``Mlpam: A machine learning and
  probabilistic analysis based model for preserving security and privacy in
  cloud environment,'' \emph{IEEE Systems Journal}, vol.~15, no.~3, pp.
  4248--4259, 2020.

\bibitem{54}
R.~Gupta and A.~K. Singh, ``Differential and access policy based
  privacy-preserving model in cloud environment,'' \emph{Journal of Web
  Engineering}, vol.~21, no.~3, pp. 609--632, 2022.

\bibitem{55}
------, ``Privacy-preserving cloud data model based on differential approach,''
  in \emph{2022 Second International Conference on Power, Control and Computing
  Technologies (ICPC2T)}.\hskip 1em plus 0.5em minus 0.4em\relax IEEE, 2022,
  pp. 1--6.

\bibitem{56}
S.~Chhabra and A.~K. Singh, ``Dynamic resource allocation method for load
  balance scheduling over cloud data center networks,'' \emph{Journal of Web
  Engineering}, vol.~20, no.~8, pp. 2269--2284, 2021.

\bibitem{57}
------, ``A secure vm allocation scheme to preserve against co-resident
  threat,'' \emph{International Journal of Web Engineering and Technology},
  vol.~15, no.~1, pp. 96--115, 2020.

\bibitem{58}
J.~Kumar, A.~K. Singh, and A.~Mohan, ``Resource-efficient load-balancing
  framework for cloud data center networks,'' \emph{ETRI Journal}, vol.~43,
  no.~1, pp. 53--63, 2021.

\bibitem{59}
R.~Gupta, D.~Saxena, and A.~K. Singh, ``Data security and privacy in cloud
  computing: concepts and emerging trends,'' \emph{arXiv preprint
  arXiv:2108.09508}, 2021.

\bibitem{60}
S.~Chhabra and A.~Singh, ``Dynamic hierarchical load balancing model for cloud
  data centre networks,'' \emph{Electronics Letters}, vol.~55, no.~2, pp.
  94--96, 2019.

\bibitem{61}
A.~K. Singh and R.~Gupta, ``A privacy-preserving model based on differential
  approach for sensitive data in cloud environment,'' \emph{Multimedia Tools
  and Applications}, vol.~81, no.~23, pp. 33\,127--33\,150, 2022.

\bibitem{62}
S.~Chhabra and A.~K. Singh, ``Optimal vm placement model for load balancing in
  cloud data centers,'' in \emph{2019 7th International Conference on Smart
  Computing \& Communications (ICSCC)}.\hskip 1em plus 0.5em minus 0.4em\relax
  IEEE, 2019, pp. 1--5.

\bibitem{63}
J.~Kumar and A.~K. Singh, ``Cloud resource demand prediction using differential
  evolution based learning,'' in \emph{2019 7th International Conference on
  Smart Computing \& Communications (ICSCC)}.\hskip 1em plus 0.5em minus
  0.4em\relax IEEE, 2019, pp. 1--5.

\bibitem{64}
------, ``Performance assessment of time series forecasting models for cloud
  datacenter networks’ workload prediction,'' \emph{Wireless Personal
  Communications}, vol. 116, no.~3, pp. 1949--1969, 2021.

\bibitem{65}
D.~Saxena and A.~K. Singh, ``Vm failure prediction based intelligent resource
  management model for cloud environments,'' in \emph{2022 Second International
  Conference on Power, Control and Computing Technologies (ICPC2T)}.\hskip 1em
  plus 0.5em minus 0.4em\relax IEEE, 2022, pp. 1--6.

\bibitem{66}
J.~Kumar and A.~K. Singh, ``Adaptive learning based prediction framework for
  cloud datacenter networks' workload anticipation.'' \emph{Journal of
  Information Science \& Engineering}, vol.~36, no.~5, 2020.

\bibitem{67}
D.~Saxena and A.~Singh, ``Security embedded dynamic resource allocation model
  for cloud data centre,'' \emph{Electronics Letters}, vol.~56, no.~20, pp.
  1062--1065, 2020.

\bibitem{68}
D.~Saxena and A.~K. Singh, ``Auto-adaptive learning-based workload forecasting
  in dynamic cloud environment,'' \emph{International Journal of Computers and
  Applications}, vol.~44, no.~6, pp. 541--551, 2022.

\bibitem{69}
------, ``Osc-mc: Online secure communication model for cloud environment,''
  \emph{IEEE Communications Letters}, vol.~25, no.~9, pp. 2844--2848, 2021.

\bibitem{70}
J.~Kumar, D.~Saxena, A.~K. Singh, and A.~Mohan, ``Biphase adaptive
  learning-based neural network model for cloud datacenter workload
  forecasting,'' \emph{Soft Computing}, vol.~24, no.~19, pp. 14\,593--14\,610,
  2020.

\bibitem{harzing2016google}
A.-W. Harzing and S.~Alakangas, ``Google scholar, scopus and the web of
  science: a longitudinal and cross-disciplinary comparison,''
  \emph{Scientometrics}, vol. 106, pp. 787--804, 2016.

\bibitem{71}
D.~Saxena, I.~Gupta, J.~Kumar, A.~K. Singh, and X.~Wen, ``A secure and
  multiobjective virtual machine placement framework for cloud data center,''
  \emph{IEEE Systems Journal}, vol.~16, no.~2, pp. 3163--3174, 2021.

\bibitem{72}
D.~Saxena and A.~K. Singh, ``Communication cost aware resource efficient load
  balancing (care-lb) framework for cloud datacenter,'' \emph{Recent Advances
  in Computer Science and Communications}, vol.~12, pp. 1--00, 2020.

\bibitem{73}
N.~K. Sharma and G.~R.~M. Reddy, ``Multi-objective energy efficient virtual
  machines allocation at the cloud data center,'' \emph{IEEE Transactions on
  Services Computing}, vol.~12, no.~1, pp. 158--171, 2016.

\bibitem{74}
D.~Saxena and A.~K. Singh, ``Ofp-tm: an online vm failure prediction and
  tolerance model towards high availability of cloud computing environments,''
  \emph{The Journal of Supercomputing}, vol.~78, no.~6, pp. 8003--8024, 2022.

\bibitem{75}
A.~K. Singh, N.~Singh, and I.~Gupta, ``Cloud-hpa: hierarchical privacy
  perseverance anatomy for data storage in cloud environment,''
  \emph{Multimedia Tools and Applications}, vol.~83, no.~13, pp.
  37\,431--37\,451, 2024.

\bibitem{76}
N.~Singh, J.~Kumar, A.~K. Singh, and A.~Mohan, ``Privacy-preserving
  multi-keyword hybrid search over encrypted data in cloud,'' \emph{Journal of
  Ambient Intelligence and Humanized Computing}, vol.~15, no.~1, pp. 261--274,
  2024.

\bibitem{77}
S.~R. Swain, A.~Parashar, A.~K. Singh, and C.~N. Lee, ``An energy efficient
  virtual machine placement scheme for intelligent resource management at cloud
  data center,'' in \emph{2023 OITS International Conference on Information
  Technology (OCIT)}.\hskip 1em plus 0.5em minus 0.4em\relax IEEE, 2023, pp.
  65--70.

\bibitem{78}
S.~Patni and A.~K. Singh, ``An optimal host allocation and load distribution
  framework using maximum likelihood in cloud environment,'' \emph{SN Computer
  Science}, vol.~4, no.~5, p. 572, 2023.

\bibitem{79}
P.~Rani and A.~K. Singh, ``An efficient and privacy-preserving data aggregation
  scheme for smart grids in cloud environment,'' \emph{SN Computer Science},
  vol.~4, no.~5, p. 540, 2023.

\bibitem{80}
R.~Gupta, D.~Saxena, and A.~K. Singh, ``A data privacy-preserving model driven
  on differential approach in cloud environment,'' in \emph{International
  Conference on Green Energy, Computing and Intelligent Technology (GEn-CITy
  2023)}, vol. 2023.\hskip 1em plus 0.5em minus 0.4em\relax IET, 2023, pp.
  109--116.

\bibitem{chadegani2013comparison}
A.~A. Chadegani, H.~Salehi, M.~M. Yunus, H.~Farhadi, M.~Fooladi, M.~Farhadi,
  and N.~A. Ebrahim, ``A comparison between two main academic literature
  collections: Web of science and scopus databases,'' \emph{arXiv preprint
  arXiv:1305.0377}, 2013.

\bibitem{81}
D.~Saxena, I.~Gupta, R.~Gupta, A.~K. Singh, and X.~Wen, ``An ai-driven vm
  threat prediction model for multi-risks analysis-based cloud cybersecurity,''
  \emph{IEEE Transactions on Systems, Man, and Cybernetics: Systems}, 2023.

\bibitem{82}
R.~Gupta, D.~Saxena, and A.~K. Singh, ``Cryptography approach for secure
  outsourced data storage in cloud environment,'' \emph{arXiv preprint
  arXiv:2306.08322}, 2023.

\bibitem{83}
I.~Gupta, D.~Saxena, A.~K. Singh, and C.-N. Lee, ``Secom: An outsourced
  cloud-based secure communication model for advanced privacy preserving data
  computing and protection,'' \emph{IEEE Systems Journal}, vol.~17, no.~4, pp.
  5130--5141, 2023.

\bibitem{84}
R.~Gupta, I.~Gupta, D.~Saxena, and A.~K. Singh, ``A differential approach and
  deep neural network based data privacy-preserving model in cloud
  environment,'' \emph{Journal of Ambient Intelligence and Humanized
  Computing}, vol.~14, no.~5, pp. 4659--4674, 2023.

\bibitem{85}
D.~Saxena, A.~K. Singh, C.-N. Lee, and R.~Buyya, ``A sustainable and secure
  load management model for green cloud data centres,'' \emph{Scientific
  Reports}, vol.~13, no.~1, p. 491, 2023.

\bibitem{86}
R.~Gupta and A.~K. Singh, ``A differential privacy-based secure data sharing
  model in cloud environment,'' in \emph{2022 IEEE 6th Conference on
  Information and Communication Technology (CICT)}.\hskip 1em plus 0.5em minus
  0.4em\relax IEEE, 2022, pp. 1--6.

\bibitem{87}
A.~Yadav, S.~Kushwaha, J.~Gupta, D.~Saxena, and A.~K. Singh, ``A survey of the
  workload forecasting methods in cloud computing,'' in \emph{Proceedings of
  3rd International Conference on Machine Learning, Advances in Computing,
  Renewable Energy and Communication: MARC 2021}.\hskip 1em plus 0.5em minus
  0.4em\relax Springer, 2022, pp. 539--547.

\bibitem{88}
D.~Saxena and A.~K. Singh, ``A high availability management model based on vm
  significance ranking and resource estimation for cloud applications,''
  \emph{IEEE Transactions on Services Computing}, vol.~16, no.~3, pp.
  1604--1615, 2022.

\bibitem{89}
S.~Chhabra and A.~K. Singh, ``A comprehensive vision on cloud computing
  environment: Emerging challenges and future research directions,''
  \emph{arXiv preprint arXiv:2207.07955}, 2022.

\bibitem{90}
D.~Saxena, R.~Gupta, A.~K. Singh, and A.~Vasilakos, ``Emerging vm threat
  prediction and dynamic workload estimation for secure resource management in
  industrial clouds,'' \emph{IEEE Transactions on Automation Science and
  Engineering}, 2023.

\bibitem{chen2019visualizing}
C.~Chen and M.~Song, ``Visualizing a field of research: A methodology of
  systematic scientometric reviews,'' \emph{PloS one}, vol.~14, no.~10, p.
  e0223994, 2019.

\bibitem{chen2012emerging}
C.~Chen, Z.~Hu, S.~Liu, and H.~Tseng, ``Emerging trends in regenerative
  medicine: a scientometric analysis in citespace,'' \emph{Expert opinion on
  biological therapy}, vol.~12, no.~5, pp. 593--608, 2012.

\bibitem{jacomy2014forceatlas2}
M.~Jacomy, T.~Venturini, S.~Heymann, and M.~Bastian, ``Forceatlas2, a
  continuous graph layout algorithm for handy network visualization designed
  for the gephi software,'' \emph{PloS one}, vol.~9, no.~6, p. e98679, 2014.

\bibitem{van2010software}
N.~Van~Eck and L.~Waltman, ``Software survey: Vosviewer, a computer program for
  bibliometric mapping,'' \emph{scientometrics}, vol.~84, no.~2, pp. 523--538,
  2010.

\bibitem{cobo2012scimat}
M.~J. Cobo, A.~G. L{\'o}pez-Herrera, E.~Herrera-Viedma, and F.~Herrera,
  ``Scimat: A new science mapping analysis software tool,'' \emph{Journal of
  the American Society for information Science and Technology}, vol.~63, no.~8,
  pp. 1609--1630, 2012.

\bibitem{johnson1987ucinet}
J.~D. Johnson, ``Ucinet: a software tool for network analysis,'' 1987.

\bibitem{chen2006citespace}
C.~Chen, ``Citespace ii: Detecting and visualizing emerging trends and
  transient patterns in scientific literature,'' \emph{Journal of the American
  Society for information Science and Technology}, vol.~57, no.~3, pp.
  359--377, 2006.

\bibitem{91}
J.~Kumar, D.~Saxena, and S.~Rekha, ``A review on sustainable resource
  management in cloud environment,'' 2023.

\bibitem{92}
D.~Saxena and S.~Rekha, ``Defensive countermeasures towards addressing cloud
  security attacks,'' 2023.

\bibitem{93}
S.~Chhabra, D.~Saxena, and S.~Rekha, ``A review on secure cloud resource
  management,'' 2023.

\bibitem{94}
S.~R. Swain, D.~Saxena, J.~Kumar, A.~K. Singh, and C.-N. Lee, ``An ai-driven
  intelligent traffic management model for 6g cloud radio access networks,''
  \emph{IEEE Wireless Communications Letters}, 2023.

\bibitem{95}
J.~Kumar, R.~Gupta, D.~Saxena, and A.~K. Singh, ``Power consumption forecast
  model using ensemble learning for smart grid,'' \emph{The Journal of
  Supercomputing}, pp. 1--22, 2023.

\bibitem{96}
J.~Kumar and A.~K. Singh, ``A discussion and comparative study on security and
  privacy of smart meter data,'' \emph{arXiv preprint arXiv:2111.09227}, 2021.

\bibitem{97}
------, ``Security and privacy-preservation of iot data in cloud-fog computing
  environment,'' \emph{arXiv preprint arXiv:2212.00321}, 2022.

\bibitem{98}
------, ``A demand and response management model using load forecasting
  technique for smart grid,'' in \emph{2023 5th International Conference on
  Energy, Power and Environment: Towards Flexible Green Energy Technologies
  (ICEPE)}.\hskip 1em plus 0.5em minus 0.4em\relax IEEE, 2023, pp. 1--6.

\bibitem{99}
D.~Saxena, K.~Gupta, R.~Gupta, J.~Kumar, and A.~K. Singh, ``Fedmup: Federated
  learning driven malicious user prediction model for secure data distribution
  in cloud environments.''

\bibitem{100}
R.~Gupta and A.~K. Singh, ``A differential approach for data and classification
  service-based privacy-preserving machine learning model in cloud
  environment,'' \emph{New Generation Computing}, vol.~40, no.~3, pp. 737--764,
  2022.

\bibitem{fang2018climate}
Y.~Fang, J.~Yin, and B.~Wu, ``Climate change and tourism: A scientometric
  analysis using citespace,'' \emph{Journal of Sustainable Tourism}, vol.~26,
  no.~1, pp. 108--126, 2018.

\bibitem{liberati2009prisma}
A.~Liberati, D.~G. Altman, J.~Tetzlaff, C.~Mulrow, P.~C. G{\o}tzsche, J.~P.
  Ioannidis, M.~Clarke, P.~J. Devereaux, J.~Kleijnen, and D.~Moher, ``The
  prisma statement for reporting systematic reviews and meta-analyses of
  studies that evaluate health care interventions: explanation and
  elaboration,'' \emph{Annals of internal medicine}, vol. 151, no.~4, pp.
  W--65, 2009.

\bibitem{ma2022visualization}
T.~Ma, Y.~Liu, and M.~Han, ``Visualization analysis of organizational
  resilience research based on citespace from 1990--2022,'' \emph{IEEE Access},
  vol.~10, pp. 65\,854--65\,872, 2022.

\bibitem{101}
R.~Gupta and A.~K. Singh, ``A privacy-preserving model for cloud data storage
  through fog computing,'' \emph{International Journal of Computer Aided
  Engineering and Technology}, vol.~17, no.~3, pp. 348--359, 2022.

\bibitem{102}
A.~Tripathi, U.~Singh, G.~Bansal, R.~Gupta, and A.~K. Singh, ``Hedcm: Human
  emotions detection and classification model from speech using cnn,'' in
  \emph{Workshop on Advances in Computational Intelligence at ISIC}, 2021.

\bibitem{103}
A.~K. Singh, S.~Chhabra, R.~Gupta, and D.~Saxena, ``A reliable client detection
  system during load balancing for multi-tenant cloud environment,'' \emph{SN
  Computer Science}, vol.~4, no.~1, p.~86, 2022.

\bibitem{104}
M.~Sharma, R.~Mittal, A.~Bharati, D.~Saxena, and A.~K. Singh, ``Emotional
  information-based hybrid recommendation system,'' in \emph{Soft Computing for
  Problem Solving: Proceedings of the SocProS 2022}.\hskip 1em plus 0.5em minus
  0.4em\relax Springer, 2023, pp. 249--267.

\bibitem{105}
S.~R. Swain, A.~K. Singh, and C.~N. Lee, ``Efficient resource management in
  cloud environment,'' \emph{arXiv preprint arXiv:2207.12085}, 2022.

\bibitem{106}
I.~Gupta, P.~K. Yadav, S.~Pareek, S.~Shakeel, and A.~K. Singh, ``Auxiliary
  informatics system: an advancement towards a smart home environment,'' 2022.

\bibitem{107}
I.~Gupta, V.~Sharma, S.~Kaur, and A.~K. Singh, ``Pca-rf: An efficient
  parkinson's disease prediction model based on random forest classification,''
  \emph{arXiv preprint arXiv:2203.11287}, 2022.

\bibitem{108}
H.~M. Gaur, A.~K. Singh, and U.~Ghanekar, ``An efficient design of scalable
  reversible multiplier with testability,'' \emph{Journal of Circuits, Systems
  and Computers}, p. 2250179, 2022.

\bibitem{109}
I.~Gupta, H.~Mittal, D.~Rikhari, and A.~K. Singh, ``Mlrm: A multiple linear
  regression based model for average temperature prediction of a day,''
  \emph{arXiv preprint arXiv:2203.05835}, 2022.

\bibitem{110}
I.~Gupta, T.~K. Madan, S.~Singh, and A.~K. Singh, ``Hisa-smfm: Historical and
  sentiment analysis based stock market forecasting model,'' \emph{arXiv
  preprint arXiv:2203.08143}, 2022.

\bibitem{111}
I.~Gupta, S.~Mittal, A.~Tiwari, P.~Agarwal, and A.~K. Singh, ``Tidf-dlpm: Term
  and inverse document frequency based data leakage prevention model,''
  \emph{arXiv preprint arXiv:2203.05367}, 2022.

\bibitem{112}
D.~Saxena and A.~K. Singh, ``An intelligent traffic entropy learning-based load
  management model for cloud networks,'' \emph{IEEE Networking Letters},
  vol.~4, no.~2, pp. 59--63, 2022.

\bibitem{113}
I.~Gupta and A.~K. Singh, ``A holistic view on data protection for sharing,
  communicating, and computing environments: Taxonomy and future directions,''
  \emph{arXiv preprint arXiv:2202.11965}, 2022.

\bibitem{114}
A.~Makkar, T.~W. Kim, A.~K. Singh, J.~Kang, and J.~H. Park, ``Secureiiot
  environment: Federated learning empowered approach for securing iiot from
  data breach,'' \emph{IEEE Transactions on Industrial Informatics}, 2022.

\bibitem{115}
A.~Acharya, H.~Prasad, V.~Kumar, I.~Gupta, and A.~K. Singh, ``Maci: Malicious
  api call identifier model to secure the host platform,'' in \emph{Proceedings
  of the Seventh International Conference on Mathematics and Computing}.\hskip
  1em plus 0.5em minus 0.4em\relax Springer, 2022, pp. 309--320.

\bibitem{116}
R.~Patel, V.~Choudhary, D.~Saxena, and A.~K. Singh, ``Lstm and nlp based
  forecasting model for stock market analysis,'' in \emph{2021 First
  International Conference on Advances in Computing and Future Communication
  Technologies (ICACFCT)}.\hskip 1em plus 0.5em minus 0.4em\relax IEEE, 2021,
  pp. 52--57.

\bibitem{117}
D.~Saxena and A.~K. Singh, ``Communication cost aware resource efficient load
  balancing (care-lb) framework for cloud datacenter,'' \emph{Recent Advances
  in Computer Science and Communications}, vol.~12, pp. 1--00, 2020.

\bibitem{118}
J.~Kumar and A.~K. Singh, ``Performance evaluation of metaheuristics algorithms
  for workload prediction in cloud environment,'' \emph{Applied Soft
  Computing}, vol. 113, p. 107895, 2021.

\bibitem{119}
D.~Saxena, R.~Gupta, and A.~K. Singh, ``A survey and comparative study on
  multi-cloud architectures: emerging issues and challenges for cloud
  federation,'' \emph{arXiv preprint arXiv:2108.12831}, 2021.

\bibitem{120}
H.~Mittal, D.~Rikhari, J.~Kumar, and A.~K. Singh, ``A study on machine learning
  approaches for player performance and match results prediction,'' \emph{arXiv
  preprint arXiv:2108.10125}, 2021.

\bibitem{121}
P.~Tiwari, S.~Mehta, N.~Sakhuja, J.~Kumar, and A.~K. Singh, ``Credit card fraud
  detection using machine learning: A study,'' \emph{arXiv preprint
  arXiv:2108.10005}, 2021.

\bibitem{122}
H.~M. Gaur, A.~K. Singh, and U.~Ghanekar, ``Testable designs of toffoli fredkin
  reversible circuits,'' \emph{arXiv preprint arXiv:2108.07448}, 2021.

\bibitem{123}
D.~Saxena and A.~K. Singh, ``Workload forecasting and resource management
  models based on machine learning for cloud computing environments,''
  \emph{arXiv preprint arXiv:2106.15112}, 2021.

\bibitem{124}
B.~Pradhan, B.~Singh, A.~Bhoria, and A.~K. Singh, ``A comparative study on
  cipher text policy attribute based encryption schemes,'' 2021.

\bibitem{125}
D.~Varshney, B.~Babukhanwala, J.~Khan, D.~Saxena, and A.~kumar Singh, ``Machine
  learning techniques for plant disease detection,'' in \emph{2021 5th
  International Conference on Trends in Electronics and Informatics
  (ICOEI)}.\hskip 1em plus 0.5em minus 0.4em\relax IEEE, 2021, pp. 1574--1581.

\bibitem{126}
R.~Patel, V.~Choudhary, D.~Saxena, and A.~K. Singh, ``Review of stock
  prediction using machine learning techniques,'' in \emph{2021 5th
  International Conference on Trends in Electronics and Informatics
  (ICOEI)}.\hskip 1em plus 0.5em minus 0.4em\relax IEEE, 2021, pp. 840--846.

\bibitem{127}
M.~Choudhary, S.~Jha, D.~Saxena, A.~K. Singh \emph{et~al.}, ``A review of fake
  news detection methods using machine learning,'' in \emph{2021 2nd
  International Conference for Emerging Technology (INCET)}.\hskip 1em plus
  0.5em minus 0.4em\relax IEEE, 2021, pp. 1--5.

\bibitem{128}
J.~Kumar, A.~K. Singh, and R.~Buyya, ``Self directed learning based workload
  forecasting model for cloud resource management,'' \emph{Information
  Sciences}, vol. 543, pp. 345--366, 2021.

\bibitem{129}
A.~Acharya, H.~Prasad, V.~Kumar, I.~Gupta, and A.~K. Singh, ``Host platform
  security and mobile agent classification: A systematic study,'' in
  \emph{Computer Networks and Inventive Communication Technologies}.\hskip 1em
  plus 0.5em minus 0.4em\relax Springer, 2021, pp. 1001--1010.

\bibitem{130}
A.~Kesharwani, A.~Nag, A.~Tiwari, I.~Gupta, B.~Sharma, and A.~K. Singh,
  ``Real-time human locator and advance home security appliances,'' in
  \emph{Evolutionary Computing and Mobile Sustainable Networks}.\hskip 1em plus
  0.5em minus 0.4em\relax Springer, 2021, pp. 37--49.

\bibitem{131}
V.~Sharma, S.~Jalwa, A.~R. Siddiqi, I.~Gupta, and A.~K. Singh, ``A lightweight
  effective randomized caesar cipher algorithm for security of data,'' in
  \emph{Evolutionary Computing and Mobile Sustainable Networks}.\hskip 1em plus
  0.5em minus 0.4em\relax Springer, 2021, pp. 411--419.

\bibitem{132}
P.~Tiwari, S.~Mehta, N.~Sakhuja, I.~Gupta, and A.~K. Singh, ``Hybrid method in
  identifying the fraud detection in the credit card,'' in \emph{Evolutionary
  Computing and Mobile Sustainable Networks}.\hskip 1em plus 0.5em minus
  0.4em\relax Springer, 2021, pp. 27--35.

\bibitem{133}
S.~Jalwa, V.~Sharma, A.~R. Siddiqi, I.~Gupta, and A.~K. Singh, ``Comprehensive
  and comparative analysis of different files using cp-abe,'' in \emph{Advances
  in Communication and Computational Technology}.\hskip 1em plus 0.5em minus
  0.4em\relax Springer, 2021, pp. 189--198.

\bibitem{134}
A.~K. Singh and I.~Gupta, ``Online information leaker identification scheme for
  secure data sharing,'' \emph{Multimedia Tools and Applications}, vol.~79,
  no.~41, pp. 31\,165--31\,182, 2020.

\bibitem{135}
J.~Kumar and A.~K. Singh, ``Decomposition based cloud resource demand
  prediction using extreme learning machines,'' \emph{Journal of Network and
  Systems Management}, vol.~28, no.~4, pp. 1775--1793, 2020.

\bibitem{136}
I.~Gupta and A.~K. Singh, ``Seli: statistical evaluation based leaker
  identification stochastic scheme for secure data sharing,'' \emph{IET
  Communications}, vol.~14, no.~20, pp. 3607--3618, 2020.

\bibitem{137}
------, ``An integrated approach for data leaker detection in cloud
  environment.'' \emph{Journal of Information Science \& Engineering}, vol.~36,
  no.~5, 2020.

\bibitem{138}
------, ``Guim-smd: guilty user identification model using summation
  matrix-based distribution,'' \emph{IET Information Security}, vol.~14, no.~6,
  pp. 773--782, 2020.

\bibitem{139}
G.~S. Hura, A.~K. Singh, and L.~S. Hoe, \emph{Advances in Communication and
  Computational Technology: Select Proceedings of ICACCT 2019}.\hskip 1em plus
  0.5em minus 0.4em\relax Springer, 2020.

\bibitem{140}
D.~Deepika, R.~Malik, S.~Kumar, R.~Gupta, and A.~K. Singh, ``A review on data
  privacy using attribute-based encryption,'' in \emph{Proceedings of the
  International Conference on Innovative Computing \& Communications (ICICC)},
  2020.

\bibitem{141}
A.~Tripathi, U.~Singh, G.~Bansal, R.~Gupta, and A.~K. Singh, ``A review on
  emotion detection and classification using speech,'' in \emph{Proceedings of
  the International Conference on Innovative Computing \& Communications
  (ICICC)}, 2020.

\bibitem{142}
A.~S. Chauhan, D.~Rani, A.~Kumar, R.~Gupta, and A.~K. Singh, ``A survey on
  privacy-preserving outsourced data on cloud with multiple data providers,''
  in \emph{Proceedings of the International Conference on Innovative Computing
  \& Communications (ICICC)}, 2020.

\bibitem{143}
I.~Gupta and A.~Singh, ``A framework for malicious agent detection in cloud
  computing environment,'' \emph{Int J Adv Sci Technol (IJAST)}, vol. 135, pp.
  49--62, 2020.

\bibitem{145}
I.~Gupta, N.~Singh, and A.~K. Singh, ``Layer-based privacy and security
  architecture for cloud data sharing,'' \emph{Journal of Communications
  Software and Systems}, vol.~15, no.~2, pp. 173--185, 2019.

\bibitem{146}
P.~Agarwal, S.~Mittal, A.~Tiwari, I.~Gupta, A.~K. Singh, and B.~Sharma,
  ``Authenticating cryptography over network in data,'' in \emph{2019
  International Conference on Intelligent Computing and Control Systems
  (ICCS)}.\hskip 1em plus 0.5em minus 0.4em\relax IEEE, 2019, pp. 632--636.

\bibitem{147}
P.~K. Yadav, S.~Pareek, S.~Shakeel, J.~Kumar, and A.~K. Singh, ``Advancements
  and security issues of iot \& cyber physical systems,'' in \emph{2019
  International Conference on Intelligent Computing and Control Systems
  (ICCS)}.\hskip 1em plus 0.5em minus 0.4em\relax IEEE, 2019, pp. 940--945.

\bibitem{148}
I.~Gupta and A.~K. Singh, ``A confidentiality preserving data leaker detection
  model for secure sharing of cloud data using integrated techniques,'' in
  \emph{2019 7th International Conference on Smart Computing \& Communications
  (ICSCC)}.\hskip 1em plus 0.5em minus 0.4em\relax IEEE, 2019, pp. 1--5.

\bibitem{149}
N.~Singh and A.~K. Singh, ``Sql-injection vulnerabilities resolving using valid
  security tool in cloud.'' \emph{Pertanika Journal of Science \& Technology},
  vol.~27, no.~1, 2019.

\bibitem{150}
S.~Chhabra and A.~K. Singh, ``A probabilistic model for finding an optimal host
  framework and load distribution in cloud environment,'' \emph{Procedia
  Computer Science}, vol. 125, pp. 683--690, 2018.

\bibitem{151}
I.~Gupta and A.~K. Singh, ``A probabilistic approach for guilty agent detection
  using bigraph after distribution of sample data,'' \emph{Procedia Computer
  Science}, vol. 125, pp. 662--668, 2018.

\bibitem{152}
S.~Chhabra and A.~K. Singh, ``Oph-lb: Optimal physical host for load balancing
  in cloud environment.'' \emph{Pertanika Journal of Science \& Technology},
  vol.~26, no.~3, 2018.

\bibitem{153}
I.~Gupta, D.~Gurnani, N.~Gupta, C.~Singla, P.~Thakral, and A.~K. Singh,
  ``Compendium of data security in cloud storage by applying hybridization of
  encryption algorithm,'' 2022.

\bibitem{154}
N.~Singh and A.~K. Singh, ``Data privacy protection mechanisms in cloud,''
  \emph{Data Science and Engineering}, vol.~3, no.~1, pp. 24--39, 2018.

\bibitem{155}
K.~Kaur, I.~Gupta, and A.~K. Singh, ``Data leakage prevention: e-mail
  protection via gateway,'' in \emph{Journal of Physics: Conference Series},
  vol. 933.\hskip 1em plus 0.5em minus 0.4em\relax IOP Publishing, 2017, p.
  012013.

\end{thebibliography}
\end{document}